\documentclass[longauth]{aa}

\usepackage{graphicx}
\usepackage{upgreek}
\usepackage{euclid}
\usepackage{diagbox}
\usepackage{adjustbox}
\usepackage{xcolor}
\usepackage{hyperref} \hypersetup{ colorlinks=true, linkcolor=blue, filecolor=magenta, urlcolor=red, citecolor=blue}
\usepackage{caption}
\usepackage[normalem]{ulem}
\newcommand{\galsimdot}{\texttt{GalSim}}
\newcommand{\galsim}{\galsimdot~}

\newcommand{\cosmosdot}{COSMOS}
\newcommand{\cosmos}{\cosmosdot~}
\newcommand{\sersicdot}{S\'ersic}
\newcommand{\sersic}{\sersicdot~}

\usepackage{txfonts}
\usepackage{comment}
\usepackage{natbib}
\definecolor{b}{rgb}{0.12, 0.46, 0.70}
\definecolor{o}{rgb}{0.9, 0.5, 0.0}
\begin{document} 

    \title{\Euclid preparation: XIII. Forecasts for galaxy morphology with the Euclid Survey using deep generative models}

\titlerunning{\Euclid Prep. XIII. Galaxy morphology with deep generative models}
   \authorrunning{H.Bretonni\`ere, M.Huertas-Company, A.Boucaud}
\author{Euclid Collaboration: H.~Bretonnière$^{1,2}$\thanks{\email{hubert.bretonniere@universite-paris-saclay.fr}}, M.~Huertas-Company$^{2,3,4,5}$, A.~Boucaud$^{2}$, F.~Lanusse$^{6}$, E.~Jullo$^{7}$, E.~Merlin$^{8}$, D.~Tuccillo$^{9}$
% ...
}

\institute{$^{1}$ Universit\'e Paris-Saclay, CNRS, Institut d'astrophysique spatiale, 91405, Orsay, France\\
$^{2}$ Universit\'e de Paris, CNRS, Astroparticule et Cosmologie, F-75013 Paris, France\\
$^{3}$ Departamento de Astrof\'{i}sica, Universidad de La Laguna, E-38206, La Laguna, Tenerife, Spain\\
$^{4}$ Instituto de Astrof\'isica de Canarias, Calle V\'ia L\'actea s/n, E-38204, San Crist\'obal de La Laguna, Tenerife, Spain\\
$^{5}$ Observatoire de Paris, PSL Research University 61, avenue de l'Observatoire, F-75014 Paris, France\\
$^{6}$ AIM, CEA, CNRS, Universit\'{e} Paris-Saclay, Universit\'{e} de Paris, F-91191 Gif-sur-Yvette, France\\
$^{7}$ Aix-Marseille Univ, CNRS, CNES, LAM, Marseille, France\\
$^{8}$ INAF-Osservatorio Astronomico di Roma, Via Frascati 33, I-00078 Monteporzio Catone, Italy\\
$^{9}$ Instituto de Astrof\'isica de Canarias (IAC); Departamento de Astrof\'isica, Universidad de La Laguna (ULL), E-38200, La Laguna, Tenerife, Spain\\
$^{10}$ Centro de Astrof\'{\i}sica da Universidade do Porto, Rua das Estrelas, 4150-762 Porto, Portugal\\
$^{11}$ Instituto de Astrof\'isica e Ci\^encias do Espa\c{c}o, Universidade do Porto, CAUP, Rua das Estrelas, PT4150-762 Porto, Portugal\\
$^{12}$ Jodrell Bank Centre for Astrophysics, School of Physics and Astronomy, University of Manchester, Oxford Road, Manchester M13 9PL, UK\\
$^{13}$ Institut de Recherche en Astrophysique et Plan\'etologie (IRAP), Universit\'e de Toulouse, CNRS, UPS, CNES, 14 Av. Edouard Belin, F-31400 Toulouse, France\\
$^{14}$ University of Lyon, UCB Lyon 1, CNRS/IN2P3, IUF, IP2I Lyon, France\\
$^{15}$ Institute of Space Sciences (ICE, CSIC), Campus UAB, Carrer de Can Magrans, s/n, 08193 Barcelona, Spain\\
$^{16}$ Institut d’Estudis Espacials de Catalunya (IEEC), Carrer Gran Capit\'a 2-4, 08034 Barcelona, Spain\\
$^{17}$ Observatoire Astronomique de Strasbourg (ObAS), Universit\'e de Strasbourg - CNRS, UMR 7550, Strasbourg, France\\
$^{18}$ European Space Agency/ESTEC, Keplerlaan 1, 2201 AZ Noordwijk, The Netherlands\\
$^{19}$ School of Physics and Astronomy, University of Nottingham, University Park, Nottingham NG7 2RD, UK\\
$^{20}$ Department of Astronomy, University of Geneva, ch. d\'Ecogia 16, CH-1290 Versoix, Switzerland\\
$^{21}$ SRON Netherlands Institute for Space Research, Landleven 12, 9747 AD, Groningen, The Netherlands\\
$^{22}$ Kapteyn Astronomical Institute, University of Groningen, PO Box 800, 9700 AV Groningen, The Netherlands\\
$^{23}$ Institute of Cosmology and Gravitation, University of Portsmouth, Portsmouth PO1 3FX, UK\\
$^{24}$ INAF-Osservatorio di Astrofisica e Scienza dello Spazio di Bologna, Via Piero Gobetti 93/3, I-40129 Bologna, Italy\\
$^{25}$ Universit\"ats-Sternwarte M\"unchen, Fakult\"at f\"ur Physik, Ludwig-Maximilians-Universit\"at M\"unchen, Scheinerstrasse 1, 81679 M\"unchen, Germany\\
$^{26}$ Max Planck Institute for Extraterrestrial Physics, Giessenbachstr. 1, D-85748 Garching, Germany\\
$^{27}$ INAF-Osservatorio Astrofisico di Torino, Via Osservatorio 20, I-10025 Pino Torinese (TO), Italy\\
$^{28}$ INFN-Sezione di Roma Tre, Via della Vasca Navale 84, I-00146, Roma, Italy\\
$^{29}$ Department of Mathematics and Physics, Roma Tre University, Via della Vasca Navale 84, I-00146 Rome, Italy\\
$^{30}$ INAF-Osservatorio Astronomico di Capodimonte, Via Moiariello 16, I-80131 Napoli, Italy\\
$^{31}$ INAF-IASF Milano, Via Alfonso Corti 12, I-20133 Milano, Italy\\
$^{32}$ Institut de F\'{i}sica d’Altes Energies (IFAE), The Barcelona Institute of Science and Technology, Campus UAB, 08193 Bellaterra (Barcelona), Spain\\
$^{33}$ Department of Physics "E. Pancini", University Federico II, Via Cinthia 6, I-80126, Napoli, Italy\\
$^{34}$ INFN section of Naples, Via Cinthia 6, I-80126, Napoli, Italy\\
$^{35}$ Dipartimento di Fisica e Astronomia “Augusto Righi” - Alma Mater Studiorum Università di Bologna, via Piero Gobetti 93/2, I-40129 Bologna, Italy\\
$^{36}$ INAF-Osservatorio Astrofisico di Arcetri, Largo E. Fermi 5, I-50125, Firenze, Italy\\
$^{37}$ Institut national de physique nucl\'eaire et de physique des particules, 3 rue Michel-Ange, 75794 Paris C\'edex 16, France\\
$^{38}$ Centre National d'Etudes Spatiales, Toulouse, France\\
$^{39}$ Institute for Astronomy, University of Edinburgh, Royal Observatory, Blackford Hill, Edinburgh EH9 3HJ, UK\\
$^{40}$ European Space Agency/ESRIN, Largo Galileo Galilei 1, 00044 Frascati, Roma, Italy\\
$^{41}$ ESAC/ESA, Camino Bajo del Castillo, s/n., Urb. Villafranca del Castillo, 28692 Villanueva de la Ca\~nada, Madrid, Spain\\
$^{42}$ Univ Lyon, Univ Claude Bernard Lyon 1, CNRS/IN2P3, IP2I Lyon, UMR 5822, F-69622, Villeurbanne, France\\
$^{43}$ Mullard Space Science Laboratory, University College London, Holmbury St Mary, Dorking, Surrey RH5 6NT, UK\\
$^{44}$ Departamento de F\'isica, Faculdade de Ci\^encias, Universidade de Lisboa, Edif\'icio C8, Campo Grande, PT1749-016 Lisboa, Portugal\\
$^{45}$ Instituto de Astrof\'isica e Ci\^encias do Espa\c{c}o, Faculdade de Ci\^encias, Universidade de Lisboa, Campo Grande, PT-1749-016 Lisboa, Portugal\\
$^{46}$ Department of Physics, Oxford University, Keble Road, Oxford OX1 3RH, UK\\
$^{47}$ INFN-Padova, Via Marzolo 8, I-35131 Padova, Italy\\
$^{48}$ INAF-Osservatorio Astronomico di Trieste, Via G. B. Tiepolo 11, I-34131 Trieste, Italy\\
$^{49}$ Aix-Marseille Univ, CNRS/IN2P3, CPPM, Marseille, France\\
$^{50}$ Istituto Nazionale di Astrofisica (INAF) - Osservatorio di Astrofisica e Scienza dello Spazio (OAS), Via Gobetti 93/3, I-40127 Bologna, Italy\\
$^{51}$ Istituto Nazionale di Fisica Nucleare, Sezione di Bologna, Via Irnerio 46, I-40126 Bologna, Italy\\
$^{52}$ INAF-Osservatorio Astronomico di Padova, Via dell'Osservatorio 5, I-35122 Padova, Italy\\
$^{53}$ Institute of Theoretical Astrophysics, University of Oslo, P.O. Box 1029 Blindern, N-0315 Oslo, Norway\\
$^{54}$ Jet Propulsion Laboratory, California Institute of Technology, 4800 Oak Grove Drive, Pasadena, CA, 91109, USA\\
$^{55}$ von Hoerner \& Sulger GmbH, Schlo{\ss}Platz 8, D-68723 Schwetzingen, Germany\\
$^{56}$ Max-Planck-Institut f\"ur Astronomie, K\"onigstuhl 17, D-69117 Heidelberg, Germany\\
$^{57}$ Institut d'Astrophysique de Paris, 98bis Boulevard Arago, F-75014, Paris, France\\
$^{58}$ Universit\'e de Gen\`eve, D\'epartement de Physique Th\'eorique and Centre for Astroparticle Physics, 24 quai Ernest-Ansermet, CH-1211 Gen\`eve 4, Switzerland\\
$^{59}$ Department of Physics and Helsinki Institute of Physics, Gustaf H\"allstr\"omin katu 2, 00014 University of Helsinki, Finland\\
$^{60}$ NOVA optical infrared instrumentation group at ASTRON, Oude Hoogeveensedijk 4, 7991PD, Dwingeloo, The Netherlands\\
$^{61}$ Argelander-Institut f\"ur Astronomie, Universit\"at Bonn, Auf dem H\"ugel 71, 53121 Bonn, Germany\\
$^{62}$ INFN-Sezione di Bologna, Viale Berti Pichat 6/2, I-40127 Bologna, Italy\\
$^{63}$ Institute for Computational Cosmology, Department of Physics, Durham University, South Road, Durham, DH1 3LE, UK\\
$^{64}$ Universit\'e C\^{o}te d'Azur, Observatoire de la C\^{o}te d'Azur, CNRS, Laboratoire Lagrange, Bd de l'Observatoire, CS 34229, 06304 Nice cedex 4, France\\
$^{65}$ University of Applied Sciences and Arts of Northwestern Switzerland, School of Engineering, 5210 Windisch, Switzerland\\
$^{66}$ California institute of Technology, 1200 E California Blvd, Pasadena, CA 91125, USA\\
$^{67}$ Observatoire de Sauverny, Ecole Polytechnique F\'ed\'erale de Lau- sanne, CH-1290 Versoix, Switzerland\\
$^{68}$ Department of Physics and Astronomy, University of Aarhus, Ny Munkegade 120, DK–8000 Aarhus C, Denmark\\
$^{69}$ Institute of Space Science, Bucharest, Ro-077125, Romania\\
$^{70}$ Max-Planck-Institut f\"ur Astrophysik, Karl-Schwarzschild Str. 1, 85741 Garching, Germany\\
$^{71}$ Dipartimento di Fisica e Astronomia “G.Galilei", Universit\'a di Padova, Via Marzolo 8, I-35131 Padova, Italy\\
$^{72}$ Centro de Investigaciones Energ\'eticas, Medioambientales y Tecnol\'ogicas (CIEMAT), Avenida Complutense 40, 28040 Madrid, Spain\\
$^{73}$ Instituto de Astrof\'isica e Ci\^encias do Espa\c{c}o, Faculdade de Ci\^encias, Universidade de Lisboa, Tapada da Ajuda, PT-1349-018 Lisboa, Portugal\\
$^{74}$ Universidad Polit\'ecnica de Cartagena, Departamento de Electr\'onica y Tecnolog\'ia de Computadoras, 30202 Cartagena, Spain\\
$^{75}$ Infrared Processing and Analysis Center, California Institute of Technology, Pasadena, CA 91125, USA\\
$^{76}$ Dipartimento di Fisica e Astronomia, Universit\'a di Bologna, Via Gobetti 93/2, I-40129 Bologna, Italy\\
$^{77}$ INFN-Sezione di Torino, Via P. Giuria 1, I-10125 Torino, Italy\\
$^{78}$ Dipartimento di Fisica, Universit\'a degli Studi di Torino, Via P. Giuria 1, I-10125 Torino, Italy\\
$^{79}$ INAF-IASF Bologna, Via Piero Gobetti 101, I-40129 Bologna, Italy\\
$^{80}$ Space Science Data Center, Italian Space Agency, via del Politecnico snc, 00133 Roma, Italy\\
$^{81}$ IFPU, Institute for Fundamental Physics of the Universe, via Beirut 2, 34151 Trieste, Italy\\
$^{82}$ SISSA, International School for Advanced Studies, Via Bonomea 265, I-34136 Trieste TS, Italy\\
$^{83}$ INFN, Sezione di Trieste, Via Valerio 2, I-34127 Trieste TS, Italy\\
$^{84}$ Dipartimento di Fisica - Sezione di Astronomia, Universit\'a di Trieste, Via Tiepolo 11, I-34131 Trieste, Italy\\
$^{85}$ INFN-Bologna, Via Irnerio 46, I-40126 Bologna, Italy\\
$^{86}$ Dipartimento di Fisica e Scienze della Terra, Universit\'a degli Studi di Ferrara, Via Giuseppe Saragat 1, I-44122 Ferrara, Italy\\
$^{87}$ INAF, Istituto di Radioastronomia, Via Piero Gobetti 101, I-40129 Bologna, Italy\\
$^{88}$ INAF-Istituto di Astrofisica e Planetologia Spaziali, via del Fosso del Cavaliere, 100, I-00100 Roma, Italy\\
$^{89}$ School of Physics, HH Wills Physics Laboratory, University of Bristol, Tyndall Avenue, Bristol, BS8 1TL, UK\\
$^{90}$ Instituto de F\'isica Te\'orica UAM-CSIC, Campus de Cantoblanco, E-28049 Madrid, Spain\\
$^{91}$ Research Program in Systems Oncology, Faculty of Medicine, University of Helsinki, Helsinki, Finland\\
$^{92}$ Department of Physics, P.O. Box 64, 00014 University of Helsinki, Finland\\
$^{93}$ Department of Physics, Lancaster University, Lancaster, LA1 4YB, UK\\
$^{94}$ Department of Physics and Astronomy, University College London, Gower Street, London WC1E 6BT, UK\\
$^{95}$ Observational Cosmology Lab, Goddard Space Flight Center, MD 20771 and SSAI, Lanham, USA\\
$^{96}$ Helsinki Institute of Physics, Gustaf H{\"a}llstr{\"o}min katu 2, University of Helsinki, Helsinki, Finland\\
$^{97}$ Centre de Calcul de l'IN2P3, 21 avenue Pierre de Coubertin F-69627 Villeurbanne Cedex, France\\
$^{98}$ Dipartimento di Fisica "Aldo Pontremoli", Universit\'a degli Studi di Milano, Via Celoria 16, I-20133 Milano, Italy\\
$^{99}$ INFN-Sezione di Milano, Via Celoria 16, I-20133 Milano, Italy\\
$^{100}$ Dipartimento di Fisica, Sapienza Universit\`a di Roma, Piazzale Aldo Moro 2, I-00185 Roma, Italy\\
$^{101}$ Sorbonne Universit{\'e}s, UPMC Univ Paris 6 et CNRS, UMR 7095, Institut d'Astrophysique de Paris, 98 bis bd Arago, 75014 Paris, France\\
$^{102}$ ICC\&CEA, Department of Physics, Durham University, South Road, DH1 3LE, UK\\
$^{103}$ INFN, Sezione di Lecce, Via per Arnesano, CP-193, I-73100, Lecce, Italy\\
$^{104}$ Department of Mathematics and Physics E. De Giorgi, University of Salento, Via per Arnesano, CP-I93, I-73100, Lecce, Italy\\
$^{105}$ Institute for Computational Science, University of Zurich, Winterthurerstrasse 190, 8057 Zurich, Switzerland\\
$^{106}$ Departamento de F\'isica, FCFM, Universidad de Chile, Blanco Encalada 2008, Santiago, Chile\\
$^{107}$ Department of Physics, P.O.Box 35 (YFL), 40014 University of Jyv\"askyl\"a, Finland\\
}
   \date{}

\abstract{We present a machine learning framework to simulate realistic galaxies for the Euclid Survey, producing more complex and realistic galaxies than the analytical simulations currently used in \Euclid. The proposed method combines a control on galaxy shape parameters offered by analytic models with realistic surface brightness distributions learned from real \textit{Hubble} Space Telescope observations by deep generative models. We simulate a galaxy field of $0.4\,\rm{deg}^2$ as it will be seen by the \Euclid visible imager VIS, and we show that galaxy structural parameters are recovered to an  accuracy similar to that  for pure analytic \sersic profiles. Based on these simulations, we estimate that the Euclid Wide Survey (EWS) will be able to resolve the internal morphological structure of galaxies down to a surface brightness of $22.5\,\rm{mag}\,\rm{arcsec}^{-2}$, and the Euclid Deep Survey (EDS) down to  $24.9\,\rm{mag}\,\rm{arcsec}^{-2}$. This corresponds to approximately $250$ million galaxies at the end of the mission and a $50\,\%$ complete sample for stellar masses above $10^{10.6}\,\rm{M}_\odot$ (resp. $10^{9.6}\,\rm{M}_\odot$) at a redshift $z\sim0.5$ for the EWS (resp. EDS). The approach presented in this work can contribute to improving the preparation of future high-precision  cosmological imaging surveys by allowing simulations to incorporate more realistic galaxies.} 


\maketitle

\section{Introduction}

The Euclid Survey ~\citep{euclid_survey} will observe $15\, 000\,\deg^2$ ($35\,\%$ of the visible sky) over six years, both in the near-infrared and in the optical at a spatial resolution approaching that of the \textit{Hubble} Space Telescope (HST). With a field of view of $0.53\,\rm{deg}^2$, compared to that of  the HST  ($0.003\,\rm{deg}^2$), it will probe the sky at a rate around $175$ times faster. It will therefore only take around five hours to observe an area equivalent to the COSMOS field \citep{cosmos_scoville}, which is still the largest contiguous area ever observed by HST and needed around $40\,\rm{days}$ of observations. In addition to the EWS at an expected nominal depth of $24.5\,\rm{mag}$ at $10\,\sigma$ for extended sources in the visible \citep{cropper}, \Euclid will also observe $40\,\rm{deg}^2$ about two magnitudes deeper (EDS). The limiting surface brightness for the EWS in the visible will be  $29.8\,\rm{mag}\,\rm{arcsec}^{-2}$. We refer the reader to Scaramella et al. (in prep.) for precise information about the \Euclid surveys and their depths.

\Euclid will produce an unprecedented amount of high spatial resolution images that will have a lasting legacy value in a variety of scientific areas, including cosmology and galaxy formation. In order to ensure that the scientific objectives are met, realistic simulations are needed for testing and calibrating algorithms. A standard approach to simulating galaxy images is through analytic \sersic models~\citep{sersic}. It is well known that galaxies can be  modelled, to a first approximation, with two \sersic functions, one for the bulge component and the other   for the disk. \sersic models have the advantage of being fully described by three parameters: the \sersic index, which controls the steepness of the profile; the effective radius, which measures a characteristic size for the galaxy; and the axis ratio, which reflects the overall shape of the galaxy. Many previous investigations have shown that \sersic models reproduce fairly well the average surface brightness distribution of galaxies (e.g. \citealp{galfit}). However, because of their simplicity, they are not well suited to describe complex galactic structure such as spiral arms, bars, clumps, or more generally asymmetric features. This is  important for the \Euclid mission, however,  since the spatial resolution of the visible detector will permit a significant number of galaxies to be resolved. Complex galaxy morphologies can have an impact in the core science of the mission since they can affect the measurement of shear for weak lensing analysis. They are also central to a variety of scientific cases in the field of galaxy formation. The \Euclid data will be particularly important to constrain the processes that shape the structures of galaxies and quench star formation, and will allow us to  study the relations between detailed morphology, environment, active galactic nuclei activity, and stellar mass, among others (e.g.~\citealp{2008ApJ...672..177L,2014ApJ...788...28V,2013MNRAS.428.1715H,2020ApJ...897..102C,2012ApJ...744..148K,2020ApJ...895..115F, chris_morpho}). Therefore, in order to quantify the possible effects of resolved structures on the image processing pipeline algorithms and to best prepare the scientific analysis of the data, it is important to produce simulations that include realistic galaxy morphologies beyond \sersic models.

In this work we investigate a novel approach based on generative models to simulate galaxies for the Euclid Survey. We first show that our method can generate realistic \Euclid galaxy fields with a   level of control of the global shapes that is similar to that of analytic profiles, but with the addition of complex morphologies. We then use the generated images to forecast the number of galaxies for which \Euclid will resolve the internal structure.

The paper proceeds as follows. In Sect. \ref{sec:datasets} we introduce the data sets used to analyse \Euclid morphological capacities and for training our models. In Sect. \ref{sec:methods} we describe the deep generative model used in this work and its training procedure. In Sect. \ref{sec:VIS} we present our results for the generation of realistic galaxies. In Sect. \ref{sec:forecast} we use the simulated galaxies to forecast the \Euclid morphological limits. We discuss the results of the paper in Sect. \ref{sec:discussion}, and conclude in Sect. \ref{sec:summary}.

\section{Data}\label{sec:datasets}
We use two data sets for this work: the Euclid Flagship galaxy catalogue (Castander et al. in prep.), hereafter the Euclid Flagship catalogue, and the \cosmos survey~\citep{cosmos_scoville}. We use the first  to simulate best the expected \Euclid data as the goal of the paper is to forecast \Euclid capacities. The second  is used to train our deep learning model so that we lean how to simulate realistic galaxies.

\subsection{Target set: Euclid Flagship catalogue}\label{sec:TU}

To quantify the performance of our model in \Euclid-like conditions and establish morphological forecasts for the mission, we used the Euclid Flagship catalogue. We accessed the catalogue through CosmoHub, a platform that allows the management and exploration of very large catalogues, best described in \cite{cosmohub} and \cite{true_universe}.

The Flagship catalogue was built using a semi-empirical halo occupation distribution (HOD) model and was intended to reproduce the global photometric and morphological properties of galaxies as well as the clustering. We refer the reader to \cite{light_cones} for more details. In order to produce a catalogue close to the real Universe, the morphological parameters, which is what we mainly use in this work, are calibrated on the CANDELS survey~\citep{candels} and 3D model fitting on the GOODS fields \citep{goods} by Welikala et al. (in prep.). Details about the catalogue production will be presented in Castander et al. (in prep.). Each simulated galaxy in the catalogue is made of two components, a bulge and a disk. The bulge component is modelled as a \sersic profile with an index varying from $n=0.3$ to $n=6$. The disk component is rendered using an exponential profile ($n=1$). The version of the Euclid Flagship catalogue used in this work contains $710\,\rm{million}$ galaxies distributed over $1200\,\rm{deg}^2$, from which we took a random subsample of $44$ million galaxies. The distributions of the main morphological parameters used in this work are presented in Fig.~\ref{fig:data hists}: the half-light radius $r_{\rm{e}}$, the axis ratio $q$, and the \sersic index $n$. We also show the apparent magnitudes of the galaxies as measured by VIS, which is the visible imager of \Euclid (Cropper et al. in prep.), as well as the redshift and the stellar mass distributions, which we use in Sect. \ref{sec:forecast} to perform our forecasts. Finally, we show the bulge-to-disk component flux fraction (hereafter bulge fraction).
    
We note here that the Euclid Flagship catalogue is a pure tabular catalogue. The procedure currently used within the Euclid Consortium to generate the galaxies is described in Sect. \ref{sec:sim}, when we compare our galaxies to the current analytic ones. Our work in this study is to use this catalogue of double \sersic profile parameters to generate the 2D images of the internally structured galaxies.

\subsection{Training set: \cosmos}

The training set is based on the \cosmos survey. \cosmos is a survey of a $2\,\deg^2$ area with the \textit{Hubble} Space Telescope Advanced Camera for Surveys (ACS) Wide Field Channel using the F814W filter. The final drizzle pixel scale is of $\ang{;;0.03}\,\rm{pixel}^{-1}$ and the limiting point source depth at $5\,\sigma$ is $27.2\,\rm{mag}$. The central wavelength of the F814W filter roughly corresponds to that of the VIS filter ($550-900\,\rm{nm}$) and the spatial resolution and depth are better than those  expected from the Euclid Survey. Therefore, the data set is well suited and is expected to be close enough to the \Euclid data, allowing us to generate mock \Euclid fields without being affected by the dependence of morphology on wavelength and without introducing undesired effects owing to extrapolations.

Our selected sample is based on the catalogue by \cite{galsim_real_gal}, which has a magnitude limit of $25.2$ and contains $87\,630$ objects. The catalogue provides, for each galaxy, the best-fit parameters of a one-component and a two-component \sersic fit by \cite{cosmos_leauthaud}, updated in 2009. In this work, we use only the one-component fitting information. In Fig.~\ref{fig:data hists} we show the distribution of the \cosmos morphological parameters of galaxies  compared to those in the Euclid Flagship catalogue. Although the distributions are similar, there are some noticeable differences which might cause a problem. The most obvious one is  the magnitude. Since \cosmos is magnitude limited, the sample does not contain as many faint galaxies as the simulation. The half-light radii of the Euclid Flagship catalogue bulge component also extend to smaller values than those in the observations. They are also generally rounder than the observed ones, but the values of axis-ratios span a similar range. The \sersic index distributions are also different because, as explained previously, the Euclid Flagship disk component  always has a \sersic index of $1$. In addition, in the \cosmos catalogue the \sersic indices of the bulge component are clipped at $n=6$ to be compatible with \galsim, which creates a noticeable spike at the edge of the distribution. The mass fraction and redshift is derived by \cite{cosmos_laigle}. As we  show in the following sections, these differences, although present, do not have a significant effect on our methodology. The most important desirable property is that simulated galaxies cover a similar range to observations. That way, the neural network used in our model is not compelled to extrapolate. This is essentially the case in the distributions shown in Fig.~\ref{fig:data hists}, except for  very small bulge components and for very faint galaxies, both of which are not expected to present significant features. We  address these points in the following sections.

In addition to the catalogue, the authors also provide $128\times 128\,\rm{pixel}$ stamps centred on each galaxy where  neighbouring galaxies have been removed. This is important for training our model on a unique galaxy per stamp. Therefore, the impact of galaxy blending in the morphology forecasts will not be studied in this work. In addition, the size of the stamps inherently limits the size of galaxies that we will be able to generate.  The radius of the stamp being $64$ pixels, every galaxy with a half-light radius larger than $\sim2\arcsecond$ will be cut by the limits of the stamp. For this reason, in this work we are limited to, and thus only consider,  galaxies smaller than $2\arcsecond$. Nevertheless, galaxies with a radius bigger than $2\arcsecond$ represent only $0.6\,\%$ of the Euclid Flagship catalogue, and thus have no major impact on our results.

\begin{figure}
    \centering
    \includegraphics[width = \linewidth]{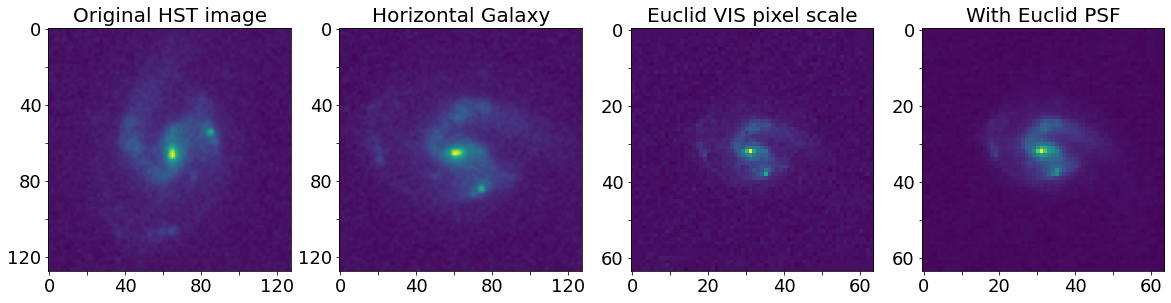}
    \caption{Illustration of our pre-processing pipeline on a random \cosmos image, and the difference between HST and \Euclid. The original image (leftmost) is rotated to be aligned with the $x$-axis of the stamp in the second image from the left, then re-scaled to the VIS resolution and cropped (third image from the left). This is the data   used to train our model. In the rightmost image the galaxy was   deconvolved by the HST PSF and re-convolved with the \Euclid PSF. This final step is shown for illustrative purposes, but is not carried out in the pre-processing of the training sample.}
    \label{fig:HST Euclid diff}
\end{figure}

The COSMOS images are pre-processed before they are used for training, as illustrated in Fig.~\ref{fig:HST Euclid diff}. We first degrade the spatial sampling from $\ang{;;0.03}\,\rm{pixel}^{-1}$ to $\ang{;;0.1}\,\rm{pixel}^{-1}$, which corresponds to the pixel scale of VIS, and then pad the image with the appropriate noise. We use the \galsim ~\citep{galsim} method described in  Sect. 5 of \citet{galsim_real_gal}. Since the pixel scale increases, the final stamp needs to be padded with noise to keep the size of $128\times128\,\rm{pixel}$. The method does this automatically by adding  a noise realisation with the same characteristics as  in the original stamps, which also takes into account the different correlations in the original noise. Doing so, the resulting images are still at the size of the \cosmos stamps. Since the pixel size is increased, we can crop up to a factor of three without losing spatial covering. However, because our model is more efficient with images that have a number of pixels which is a power of two (for parity reasons between the compression and decompression steps of our deep learning network), we crop our image by only a factor of two, resulting in images of $64\times64\,\rm{pixel}$. The purpose of this cropping is to accelerate the training. 
We finally rotate the stamps so that the galaxy semi-major axis is aligned with the $x$-axis of the image. With this  configuration we ensure that our model will learn to produce only `horizontal' galaxies and therefore position angles can be manually added in post-processing. This has the additional advantage of reducing the complexity and hence allowing the neural network to focus the attention on the more important physical properties of the object. Figure~\ref{fig:HST Euclid diff} illustrates these pre-processing steps used for the training of our model, and the final galaxy as it would be seen by VIS. Because galaxies produced by our model will be noise-free and not convolved by the PSF, we do not need to change the noise level and the PSF for the training. Thus, the inputs of our model have the noise characteristics and the PSF of the HST images. These two transformations, to go from HST to \Euclid data will be added a posteriori. More information about those transformations are described Sects. \ref{sec:sim} and \ref{sec:large_field}.

\begin{figure}
    \centering
    \includegraphics[width = \linewidth]{cosmos_tu_hists.png}
    \caption{Distributions of the main structural parameters in the data sets used in this work, along with the redshift and the stellar mass used for our forecasts. We also show the bulge to disk flux fraction (bulge fraction) for the Flagship. The $y$ axis is the normalised density counts such that the area over the curve is equal to one. For the magnitude, the COSMOS histogram shows the F814W magnitude and the Flagship one corresponds to the Euclid VIS magnitude. The range of the training set (\cosmosdot) covers most of the \Euclid data.}
    \label{fig:data hists}
\end{figure}

We use the \cosmos catalogue and images only for the training of our model. To test the performance of our model (Sect. \ref{sec:VIS}) and the forecasts (Sect. \ref{sec:forecast}), we only use the Euclid Flagship catalogue described in the previous section.

\section{\Euclid emulator with generative models}
\label{sec:methods}

In this section we describe the methodology for emulating \Euclid galaxies using the COSMOS sample described in the previous section.

The generation of synthetic data (images, language, videos) has significantly improved in  recent years thanks to new deep learning-based generative models. Generative models are a type of unsupervised machine learning algorithms that are trained to generate unseen data. There are several architectures; variational autoencoders (VAE: \citealp{VAE}), generative adversarial networks (GANs: \citealp{GAN, WGAN}), and  autoregressive models \citep{pixelcnn} are the main ones.  They all learn a probability distribution function of the pixel distribution,  which  can be sampled to generate new data. Generative models have already been used in astrophysics for a variety of different purposes. For example, with VAEs  radio galaxies can be simulated \citep{radiogals} or images of overlapping galaxies can be reconstructed separately \citep{deblending}. Using GANs, \cite{inpainting} have simulated missing data from the cosmological microwave background, while \cite{gaz_simu} have simulated gas density maps. \cite{storey_gans} and \cite{margalef_gans} have used GANs to detect outliers in imaging surveys. Autoregressive flows can be used to compare simulations and observations (e.g. \citealp{zanisi}).

In this work we use a VAE.  Variational autoencoders estimate an explicit latent space, which is an important advantage for simulating galaxies with known parameters. The compression--decompression architecture inherent to the VAEs along with the Kullback--Leibler term in the loss (see Sect. \ref{sec:fvae_methods} Eq. \ref{eq:kl}) force the latent representation to be meaningful and regular. In addition, VAEs are known to be more stable during training, and less subject to mode collapse (lack of diversity in the generation) than GANs.

\subsection{Model}
\label{subsec:emulator}

Our model for generating galaxies is based on the work by \citet{FVAE} (hereafter L2020) who describe in detail the architecture and specifics of the training procedure. We also illustrate the architecture of the two components of our model in Figs. \ref{fig:vae_architecture} and \ref{fig:flow_architecture}.

The goal of our work is to simulate and test galaxies with more realistic shapes than the classical analytic profiles while keeping a control on the shape parameters, such as axis ratios, effective radii, and fluxes. To this end, our model is made of two distinct parts: a variational  autoencoder \citep{VAE}, which learns how to simulate real galaxies from observations, and a normalising flow \citep{flow} in charge of mapping catalogue parameters to the VAE latent space. Both parts are merged together after training, resulting in an architecture called a flow variational autoencoder (FVAE). We describe in the following the global properties of these two models.

\subsubsection{Galaxy generation with a variational autoencoder}\label{sec:fvae_methods}
A VAE is a deep generative model which is trained to generate new data (galaxies) by learning a probability distribution from the training data. To this end, the VAE first compresses the input image $x$ into a low-dimensional space, also called latent space, which contains a compact and meaningful representation of the input data. Similar objects are compressed into neighbouring vectors. This is achieved with a convolutional neural network called the encoder, which can be represented as a non-linear function $E_{\vec{\Theta}}$, $\vec{\Theta}$ being its trainable parameters. While a classical autoencoder compresses the input image only into a vector $\vec{z}$, a VAE replaces that low-dimensional vector with a probability distribution function (PDF) $p_{\vec{\Theta}}(\vec{z}\,|\,x)$. In our case, $p_{\vec{\Theta}}(\vec{z}\,|\,x)$ is set to be a multivariate Gaussian distribution. This is equivalent to choosing the prior for the distribution of points in the latent space to be Gaussian. Similar galaxies will be encoded into similar regions of the distribution. Having a distribution instead of a point estimate makes the latent space continuous, allowing one to sample new regions from it and to produce new galaxies arising from the same probability density function as the data. 

A sample $\vec{z}$ is then drawn from the distribution $p_{\vec{\Theta}}$. This constitutes the input of a second convolutional neural network called the decoder $D_{\vec{\Theta'}}$, which typically has an architecture symmetric to that of the encoder. The decoder decompresses the latent representation $\vec{z}$ using transposed convolutions to produce a new image $\hat{x}$, $D_{\vec{\Theta'}}(\vec{z})=\hat{x}$. The output of the decoder can be seen as the probability that the input data $x$ effectively come from the latent space vector $\vec{z}$ (i.e. $D_{\vec{\Theta'}}(\vec{z})=p_{\vec{\Theta'}}(x\,|\,\vec{z})$). During training, the goal is to reconstruct $x$ with the best possible accuracy (i.e. $\hat{x}=x$) ensuring that the distribution encoded within the latent space is a good representation of the data.
The amount of information loss in the compression--decompression is the first term of the neural network loss function $\mathcal{L}$, which is used to adapt $\vec{\Theta}$ and $\vec{\Theta'}$ through a gradient descent minimisation. From a statistical point of view, this accuracy is defined as the negative log-likelihood of $x$ given $\vec{z}$, which can be written using the expectation value:
\begin{equation}
    \mathcal{L} = -\mathop{\mathbb{E}_{\,\vec{z}\sim p_{\vec{\Theta}}(\vec{z}\,|\,x)}\left[\log p_{\vec{\Theta '}}(x\,|\,\vec{z})\right].}
\end{equation}

In practice, we can simply see the reconstruction accuracy as the mean square error between the reconstructed image and the input:

\begin{equation}
    \mathcal{L} = \| x- \hat{x} \|^2.
\end{equation}
In addition, in order to regularise $p_{\vec{\Theta}}$, a second term is added to penalise the encoder when it produces distributions too far from a normal Gaussian distribution $\mathcal{N}(0, 1)$. This difference between $p_{\vec{\Theta}}(\vec{z}\,|\,x)$ and $\mathcal{N}(0,1)$ is estimated using the Kullback--Leibler divergence \citep{KL}:
\begin{equation}
    \mathbb{KL} = \mathbb{E}[\log p_{\vec{\Theta}}(\vec{z}\,|\,x)-\log \mathcal{N}(0,1)]\,.
    \label{eq:kl}
\end{equation}
The final loss function for the VAE reads

\begin{align}
    \mathcal{L} = - & \mathbb{E}_{\,\vec{z}\sim p_{\vec{\Theta}'}(\vec{z}\,|\,x)}\left[\log p_{\vec{\Theta} '}(x\,|\,z)\right] + \label{eq:vae loss} \nonumber \\
     & \beta \, \mathbb{E}[\log p_{\vec{\Theta}}(\vec{z}\,|\,x)\, - \log \mathcal{N}(0,1) ] \, ,
\end{align}
where $\beta$ allows us to vary the importance of the terms during training.

Lanusse et al. also introduce two additional features in order to produce images deconvolved by the PSF and without noise. To learn noise-free galaxies, a different version of the log-likelihood for the reconstruction term of the loss function is used. Instead of applying it directly to the pixels, it is done in Fourier space in order to weight the reconstruction error  less  on the high frequencies (noisy regions). The Fourier transform of the input and of the output is computed, and divided by the power spectrum of the noise. By dividing the Fourier transform of the image by the power spectrum of the noise, a smaller weight is given to the pixels with a high frequency. It ensures that the decoder learns that producing images without noise is not an error. In order to produce deconvolved images, the last convolutional layer of the decoder is not trainable and is set to be equal to the PSF. That way, the model produces an image that looks like the input image before being convolved by the PSF in the second last layer. 

\subsubsection{Sample of the shape parameters with the regressive flow}
The VAE described in the previous subsection can generate realistic galaxies by sampling from the encoded latent space. However, it cannot do so for a given size or ellipticity because it lacks the information about the mapping between the structural parameter space of the galaxy and the latent space.

To learn that mapping, L2020 propose a conditional normalising flow, based on autoregressive algorithms (MAF: \citealp{MAF}, MADE: \citealp{MADE}). A normalising flow is a bijector $g_{\vec{\Theta}}$, which transforms a distribution $q$ into another distribution $p$ with an invertible transformation $g$. We use it here to learn the mapping between a latent space with a fixed distribution $q$, referred to as the flow latent space, and the distribution $p$ inside the VAE latent space. This mapping can be made conditional to some input parameters $\vec{y}$ such as galaxy size or ellipticity. In other words, $g_{\vec{\Theta}}$ is a function of both the latent space vector $\vec{z}$ and the physical parameters of the galaxy $\vec{y}$.
 
If the mapping is well learnt, when we sample a vector $\vec{z}_{\rm{flow}}$ from the flow latent space distribution $q$ and pass it through $g_{\vec{\Theta}}$ along with a vector of physical parameter $\vec{y}$, it will output a vector $\vec{\hat{z}}$ in the VAE latent space
 \begin{equation}
\vec{\hat{z}} = g_{\vec{\Theta}}(\vec{z}_{\rm{flow}},\, \vec{y})\, , 
\end{equation}
such that $\vec{\hat{z}}$ is very similar to the vector $\vec{z}$, which would have been encoded by the VAE's encoder from a galaxy $x$ with physical parameters $\vec{y}$
\begin{equation}
\vec{\hat{z}} \approx E_{\vec{\Theta}}(x_{\vec{y}})\,.
\end{equation}
With this mapping, we now know where to sample into the VAE latent space in order to decode a galaxy with precise physical parameters: to simulate a galaxy, we need to map $\vec{z}_{\rm{flow}}$ and $\vec{y}$ to the VAE latent space, and then decode the vector with the decoder to produce an image of a galaxy that has the physical properties given by $\vec{y}$.

In practice, the training procedure is done the other way around: we learn how to map a vector $\vec{z} = E_{\vec{\Theta}}(x)$ into a vector $\vec{z}_{\rm{flow}}$ of the flow latent space. Because $g_{\vec{\Theta}}$ is a bijector, learning the mapping from the flow latent space to the VAE latent space or the other way around is the same task, but doing it in this direction is much easier because of the loss. The loss we use is the negative log likelihood of $\vec{z}$ under the distribution of the flow latent space $q$
\begin{align}
\mathcal{L}_{\rm{flow}} & = \mathbb{E}_{\vec{z}\sim{p}}\left[-\log p(z) \right ] \nonumber \\
& = \mathbb{E}_{\vec{z}\sim{p}}\left[-\log \,q\left(g^{-1}(\vec{z})\right) + \log \rm{det}\,J_{g^{-1}}(\vec{z}) \right ] ,\\
& = \mathbb{E}_{\vec{z}_{\rm{flow}}\sim{q}}\left[-\log q\,(\vec{z}_{\rm{flow}}) + \log \rm{det}\,J_{g}(\vec{z}_{\rm{flow}}) \right ],
\end{align}
where $\rm{det}\,J_g$, the determinant Jacobian of $g$, comes from the transformation between the two distributions.

Choosing a standard Gaussian distribution for $q$, we ensure that this loss is tractable (i.e. easy to compute). By construction, the Jacobian of $g$ is also easy to compute \citep{normalizing_flows}. Thus, during training, every galaxy $x$ is encoded by the previously trained encoder $E$ into a vector $\vec{z}$ drawn from the encoded distribution $p_{\vec{\Theta}}(\vec{z}\,|\,x)$. This vector $\vec{z}$ is transformed by the flow's bijector $g_{\vec{\Theta}}^{-1}$ into a vector $\vec{z}_{\rm{flow}}$ conditioned by the physical parameters of the galaxy $\vec{y}$
\begin{equation}
\vec{z}_{\rm{flow}} = g_{\vec{\Theta}}^{-1}(\vec{z},\,\vec{y})\, , 
\end{equation}
which is used to compute the loss and optimise the weights of $g$.

To implement the flow, we use the probabilistic library of \texttt{TensorFlow}, \texttt{TensorFlow probability}. With this library it becomes straightforward to implement the bijector $g$, with a chain of masked autoregressive layers, described in \cite{MADE}. The transformations of the distribution made by the successive layers (shifts of the mean and stretch of the dispersion) are conditioned to the physical parameters of the flow's input. Then, thanks to the \texttt{Distribution} object of the library, with only one command it is possible  to sample the transformed distribution (e.g. to get $\vec{z}_{\rm{flow}}$), but also to take the log likelihood for the computation of the loss.

\subsubsection{Final model}
\begin{figure}
    \centering
    \includegraphics[width = \linewidth]{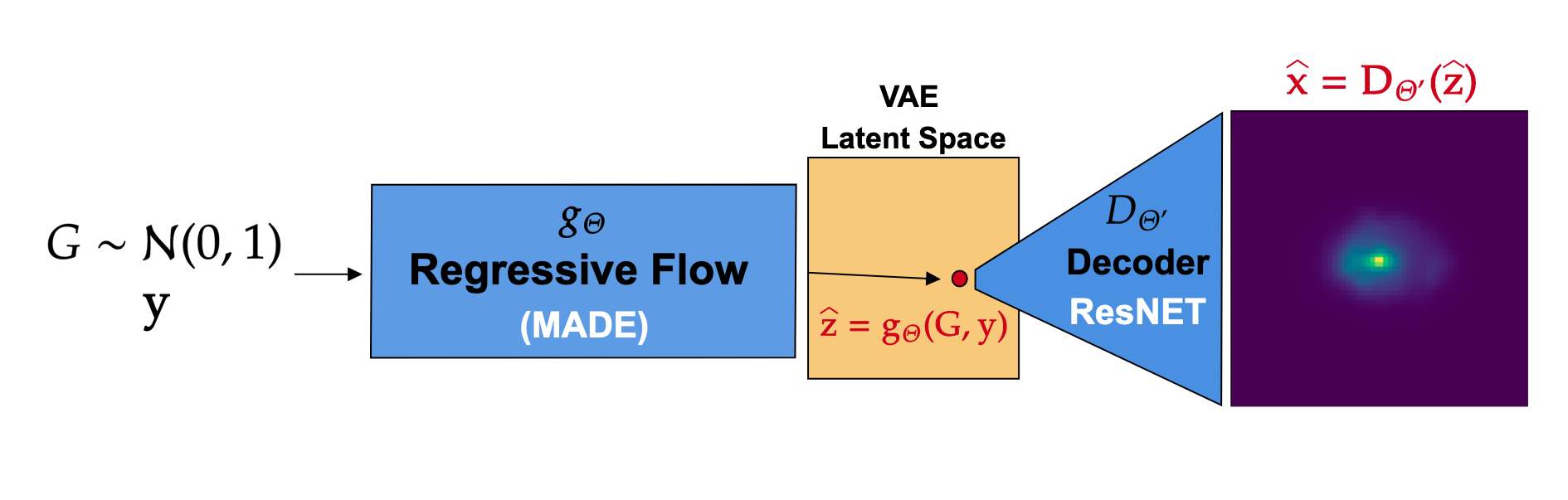}
    \caption{Schematic representation of the FVAE architecture used to simulate a galaxy with structural parameters $y$. A random noise $G$ is passed through a regressive flow conditioned to the input galaxy parameters $y$. The flow outputs a latent space vector $\hat{z}$, which is decoded by the VAE in order to produce a galaxy corresponding to the input shape parameters.}
    \label{fig:FVAE}
\end{figure}

The final model (schematic representation in Fig.~\ref{fig:FVAE}) combines the decoder part of the VAE with the regressive flow described in the previous subsection. Therefore, the input of the final model is a galaxy catalogue. The flow samples a Gaussian noise vector, which is concatenated with the catalogue parameters to produce a vector in the latent space. The vector is then decoded by the generator of the VAE, producing the image of a new galaxy with the corresponding input parameters from the catalogue. The use of a continuous distribution enables the generation of new galaxies that resemble real ones, but have never been observed before.

\subsection{Training procedure}

The main goal of this work is to produce \Euclid-like realistic galaxies. We use pre-processed \cosmos galaxies (described in Sect. \ref{sec:datasets}) to train the VAE. We train it for $250\,000$ steps, which means $3900\,\rm{epochs}$ (one epoch is when the whole training set has been seen by the network) with a batch size of $64$ (the batch size is the number of images with which we perform each gradient descent). The latent space has a dimensionality of $32$.  The learning rate has a first phase where it linearly increases, followed by a square root decay. We use a warm-up phase of $30\,\rm{epochs}$ where we train only the generative part ($\beta=0$ in Eq.~\ref{eq:vae loss}), and then linearly increase it to have the same weight between the generative term of the loss function and the $\mathbb{KL}$ ($\beta=1$). Training and validation losses converge long before the end of training. However, even after the convergence, we still see a significant improvement in the generated images. The model first learns the global shape of the galaxies and a Gaussian posterior in the latent space, making the objective function Eq.~(\ref{eq:vae loss}) already very low. The learning of more complex structures inside the galaxies does not have a great impact on the loss (most of the galaxies do not present major structures and the pixels belonging to the structures represent a small fraction of the image), which can explain why we need to train longer than the convergence to learn the complex distribution of the training set. We  show in the following sections that we chose an appropriate number of epochs to produce complex galaxies without overfitting. We did not try to optimise this number of epochs, the balance between results and training time being sufficient for our study. Nevertheless, the large number of epochs is not unusual, and generative models such as VAEs usually require a large number of epochs to converge.

In a second step, we tackle the regressive flow. We condition the model with three parameters: \sersic index $n$, half-light radius $r_{\rm{e}}$, and axis ratio $q$. We trained it for $470$ epochs, ensuring that both our training and validation loss had converged. We use a batch size of $128$, and the same learning rate strategy as for the VAE. By design, the dimensionality of the flow latent space is the same as that of the VAE (i.e. $32$ in this work).

\section{Emulation of VIS images}\label{sec:VIS}
In this section we analyse the properties of simulated galaxies and assess the accuracy of the emulation. Our emulator is expected to fulfil two main goals: realistic galaxies and  a control on the global shape parameters.

\subsection{Simulation of composite galaxies}
\subsubsection{Simulations with pure \sersic profiles}\label{sec:sim}
The Euclid Consortium currently creates analytic galaxies with the \galsim software \citep{galsim}. Each galaxy is created as the sum of two components, the bulge and the disk. The disk component is created with an exponential profile (\sersic profile with $n=1$). The bulge component is a 3D \sersic profile, which is projected to produce the expected ellipticity. The two profiles are created with the expected bulge-to-disk flux fraction, and then summed pixel-wise. The flux is then rescaled to match the total galaxy magnitude. The image is finally convolved with the VIS PSF, which has a full width at half maximum (FWHM) of $\ang{;;0.17}$ at $800\,\rm{nm}$ \citep{psf}. This PSF takes into account all the optical and instrumental effects, and thus goes beyond a simple Gaussian. It is the result of the detailed analysis of the VIS instrument performed by the Euclid Consortium. If necessary, we also rotate the galaxy to its corresponding position angle in the sky. At this stage, the galaxies are noise-free. The method used to add noise is explained in Sect. \ref{sec:large_field}.

\subsubsection{Simulations with the FVAE}
Once trained, our model takes as input the three shape parameters of each component of the galaxy from the Euclid Flagship catalogue (half-light radius $r_{\rm{e}}$, \sersic index $n$, and axis ratio $q$) and generates a galaxy with the expected structure and realistic morphology. As explained above, galaxies in the Flagship catalogue are described by two components, a bulge and a disk. To simulate exactly the same field and compare to the current \Euclid simulations, we also need to produce the two components separately. This way, we can reproduce the same method as the current Euclid procedure explained in the previous subsection. Each component (bulge and disk) is simulated separately by our model, and then added with the appropriate bulge-to-disk flux ratio. We then use \galsim to scale the flux, to convolve by the PSF, and to rotate the galaxy to the appropriate position angle. Since the flux is calibrated in the post-processing step, we can associate faint magnitudes with our emulation even if not properly covered by our training set, as shown in Sect. \ref{sec:datasets}. For the other parameters, as the distributions of the bulges and the disks in the Flagship are covered by the training set, simulating the two components separately should not be an issue.

\subsection{Qualitative inspection}
\subsubsection{Individual noise-free galaxy simulation}

\begin{figure}
    \centering
    \includegraphics[width = \linewidth]{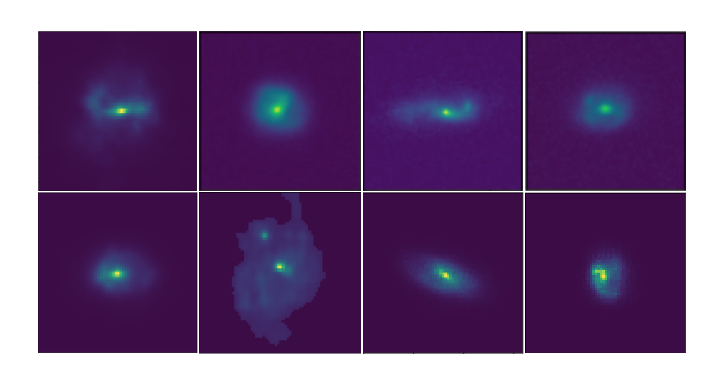}
    \caption{Example of galaxies simulated by the FVAE presenting obvious complexity and features. The scale is linear.}
    \label{fig:structures}
\end{figure}

We first qualitatively evaluate our simulations. Figure~\ref{fig:structures} shows eight galaxies with large radius, prone to presenting interesting morphologies. Compared to pure \sersic profile simulations, the generated galaxies are more complex and asymmetric (see    Fig.~\ref{fig:sersic_gals} for some examples of pure \sersic galaxies). We are able to generate the commonly observed features such as rings, spiral arms, irregularities, and clumps with different inclination angles. This visual inspection is a first indication that we are able to generate complex behaviour and mimic surface brightness profiles or features superior to those of \sersic profile simulations.

\begin{figure}
    \centering
    \includegraphics[width = \linewidth]{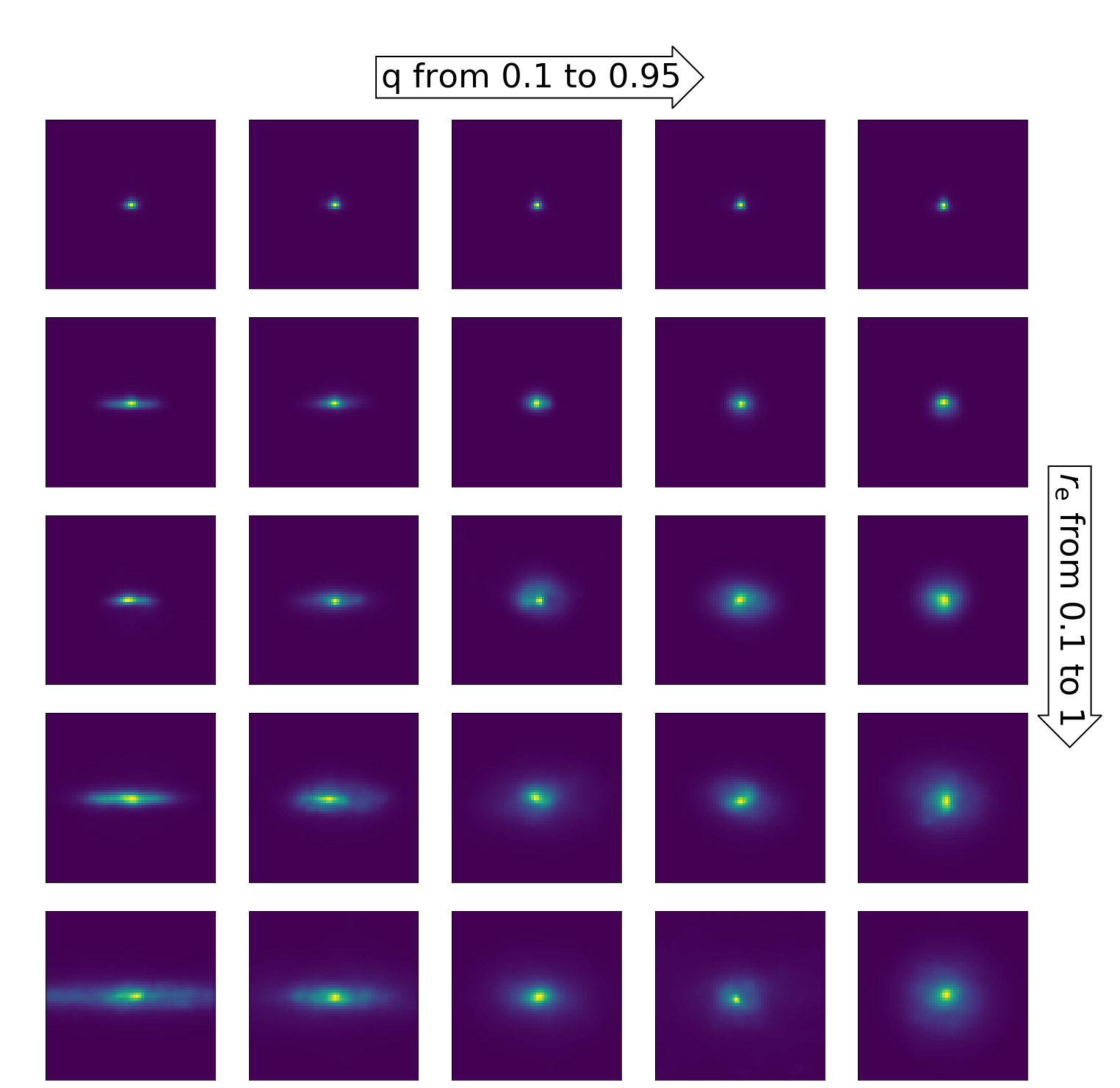}
    \caption{Galaxies simulated by our model from a catalogue with increasing axis ratios ($q$) and effective radius ($r_{\rm{e}}$). The magnitude and the \sersic index are fixed to $24$ and $1,$ respectively,  for all galaxies. The images are all $64\times 64$ pixel, the natural output of our model. Each row shows galaxies with constant $r_{\rm{e}}$, and linearly increasing $q$ from $0.1$ to $0.95$. Each column shows galaxies with fixed $q$, and linearly increasing $r_{\rm{e}}$ from $\ang{;;0.1}$ to $1\arcsecond$. The galaxies are clearly rounder and bigger from left to right and top to bottom.}
    \label{fig:q_vs_hlr}
\end{figure}

\begin{figure}
    \centering
    \includegraphics[width = \linewidth]{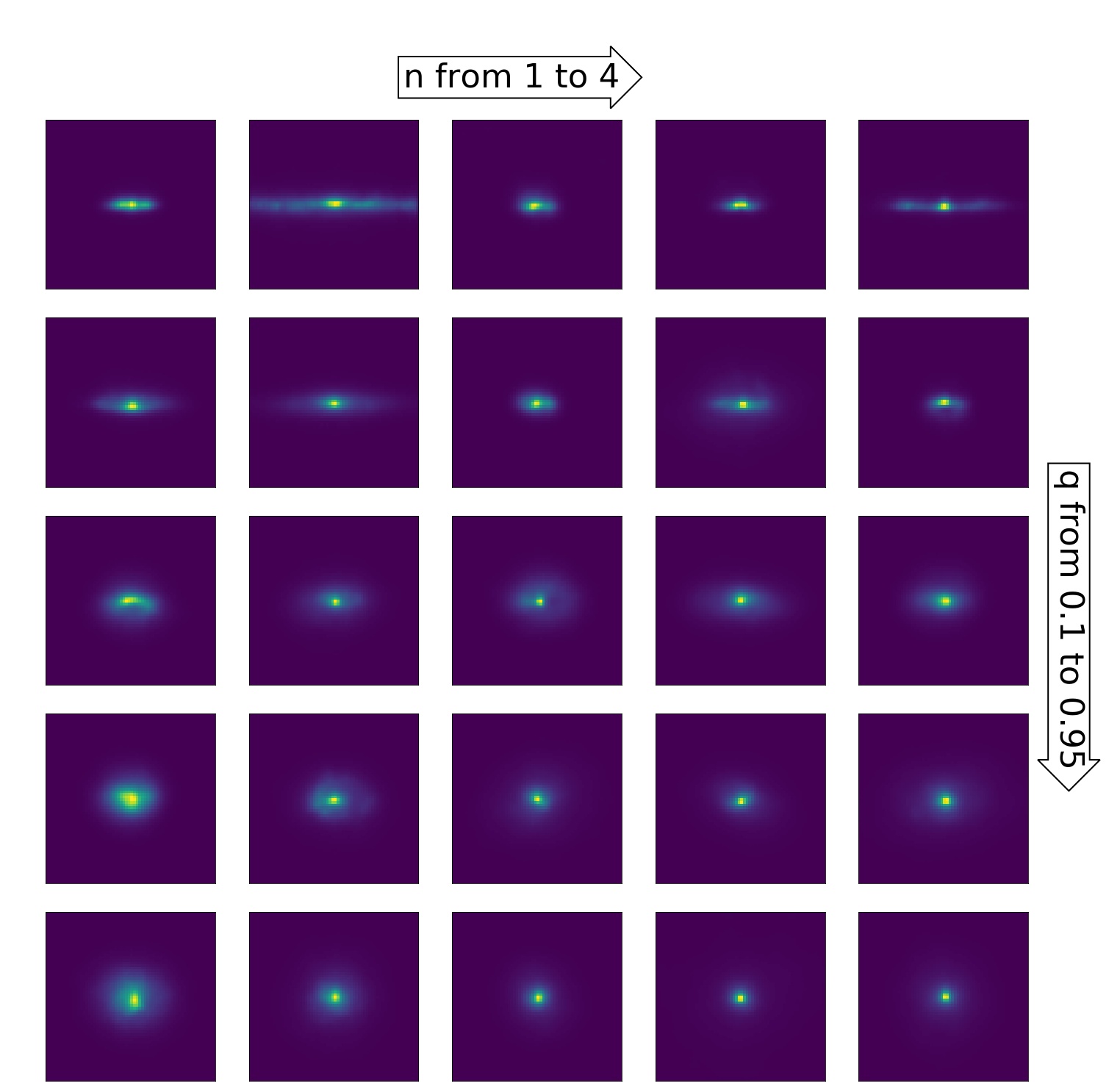}
    \caption{Galaxies simulated by our model from a catalogue with increasing axis ratios ($q$) and \sersic indices ($n$). The magnitude and the effective radius  are fixed  to $24$ and $\ang{;;0.7}$, respectively, for all galaxies. Each row shows galaxies with constant $q$, and linearly increasing $n$ from $1$ to $4$. Each column shows galaxies with fixed $n$, and linearly increasing $q$ from $0.1$ to $0.95$. The galaxies   clearly show a steeper profile and are rounder from left to right and top to bottom, respectively.}
    \label{fig:q_vs_n}
\end{figure}

The second key element of our emulator is the ability to control the structural parameters. In order to illustrate this, we show in Figs. \ref{fig:q_vs_hlr} and \ref{fig:q_vs_n} the impact of varying parameters on the generated galaxies. Figure~\ref{fig:q_vs_hlr} shows a series of generated galaxies with a constant magnitude set to $24$, a fixed \sersic index of $1.5$ and a varying axis-ratio $q$ and half-light radius $r_{\rm{e}}$. Figure~\ref{fig:q_vs_n} shows a grid of galaxies with fixed $r_{\rm{e}}$ and magnitude but varying axis-ratio and \sersic index. We can clearly observe the expected trends. Galaxies become rounder as we move from left to right, and bigger from top to bottom in Fig~\ref{fig:q_vs_hlr}. In Fig.~\ref{fig:q_vs_n} galaxies become more concentrated as the \sersic index increases from left to right.  The images also show several examples presenting non-trivial symmetric shapes. An important limitation to note is that our model is fixed to produce images of size $64\times64$ pixel. Very large galaxies might therefore be truncated.

\subsubsection{Large field simulation}\label{sec:large_field}
\begin{figure*}
    \centering
    \includegraphics[scale=0.18]{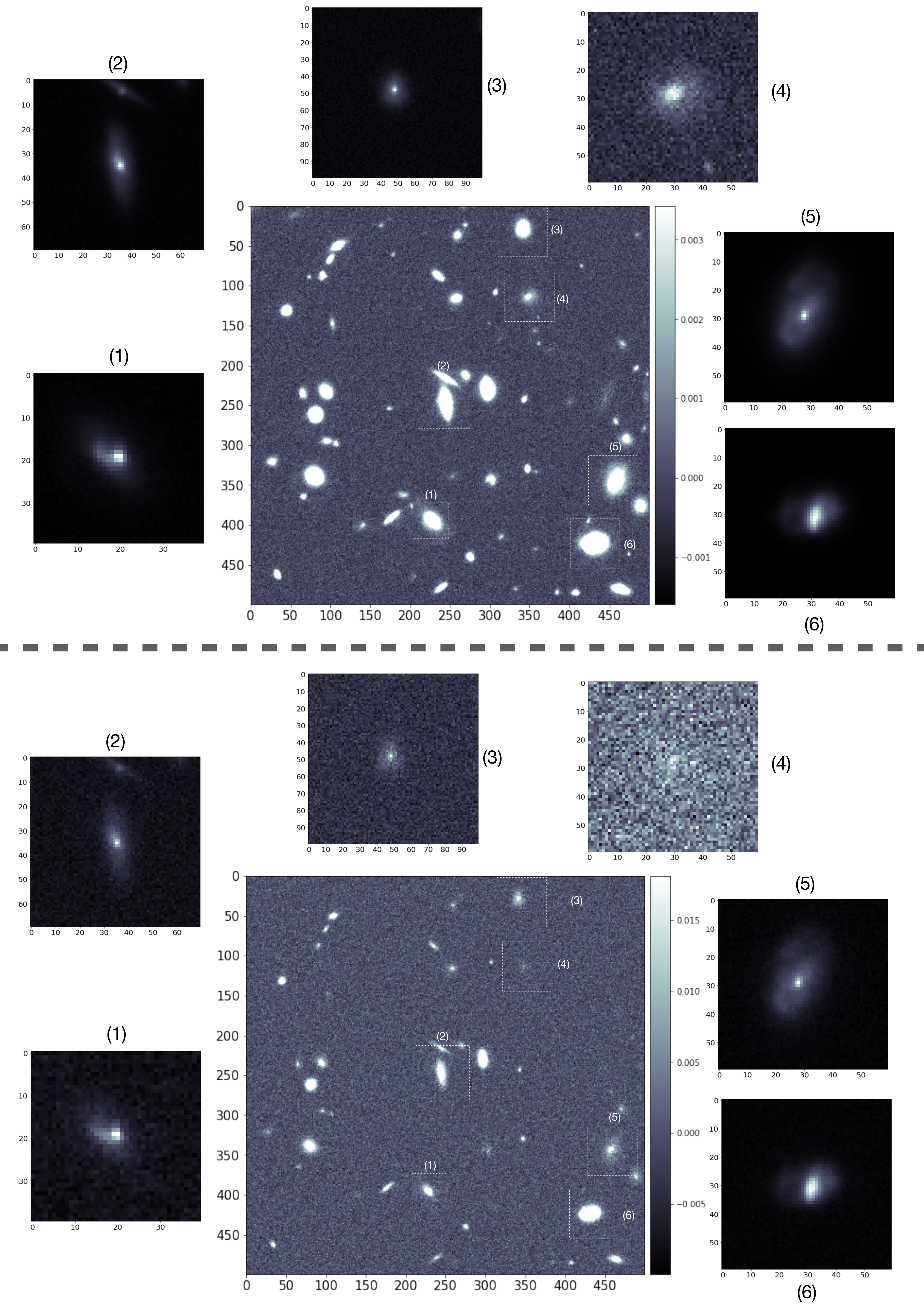}
    \caption{Illustration of a large field simulation produced by our FVAE. The top and bottom panels show the same field simulated at the depths of the EDS and the EWS, respectively. The stamps show zoomed-in regions where some galaxies present morphological diversity. In the large field images, we use the \texttt{IRAF} `zscale' that stretches and clips the low and high values to better highlight  the differences between the EWS and the EDS. The stamps are in linear scale, which better emphasises the structures. With the stretching induced by the zscale, all the structures disappear and only the global shape is still recognisable. Finally, the apparent different level of background between the stamps and the global image is also due to the different brightness scale (different maximum and minimum values in each of them).}
    \label{fig:wide and deep}
\end{figure*}

In addition to individual stamps, we also generate two large fields of $0.4\,\rm{deg}^2$ at the depths of the EWS and the EDS (see a portion of those fields in Fig.~\ref{fig:wide and deep}). We take a  subsample of the Euclid Flagship catalogue and generate every galaxy without noise and deconvolved by the PSF. We then convolve the stamp by a unique VIS PSF (no PSF variations are modelled). All the stamps are then placed in the large field into their corresponding positions according to the catalogue. We finally add the expected noise level of the EWS and the EDS in two different realisations of the same field. The background noise (coming mostly from background sources and from the zodiacal light) is simulated by  Gaussian noise with the expected standard deviation for the VIS camera (Cropper et al. in prep.; Scaramella et al. in prep.; priv. comm.). The photon noise is simulated with a Poisson distribution added to every pixel, considering the cumulative exposure times presented by \cite{euclid_survey}.

More information will be given about the noise realisations in Merlin et al. (in prep.). We do not simulate any instrumental effects such as cosmic rays, ghosts, charge transfer inefficiency, or read-out noise, considering thus an ideal case of a VIS image processing pipeline. In Fig.~\ref{fig:wide and deep} we show a random region of the large fields, and highlight some interesting galaxies.

\subsection{Quantification of structural properties}\label{sec:quantif}
This visual assessment of the previous subsection confirms that our model behaves as expected both in generating complex shapes and controlling structural parameters. However, in order for the simulation to be useful to test algorithms, it is required that the control on the structural parameters is comparable to what is achieved with analytic profiles.

\subsubsection{Surface brightness profiles}\label{sec:profiles}

\begin{figure*}[ht]
    \centering
    \includegraphics[width=\linewidth]{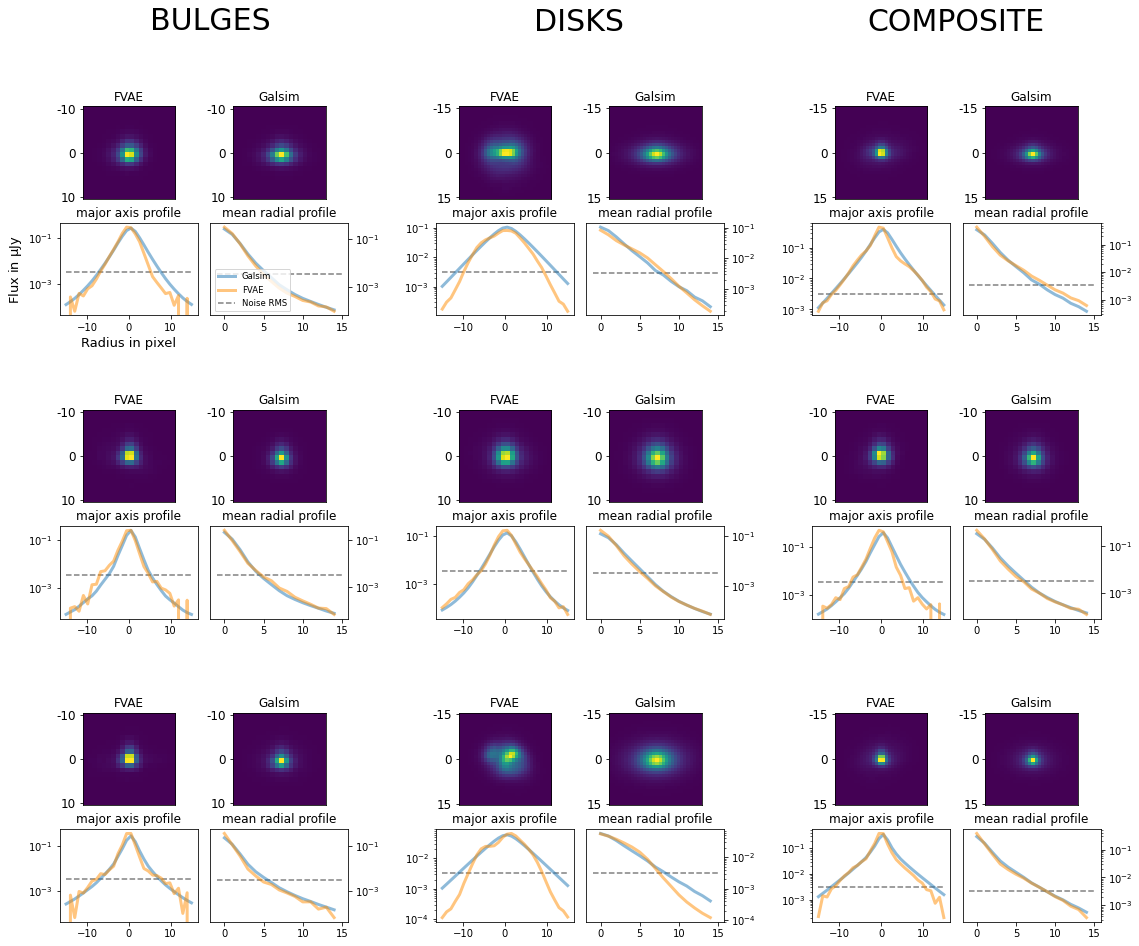}
    \caption{Examples of three radial profiles of galaxies generated with \galsim and our model. Each group of two columns represents the different components of the galaxy: bulge, disk, and composite (bulge plus disk) from left to right. Within each group the top row shows the images by our model (left) and by the \sersic model (right). The bottom line represents the light radial profiles, along the major axis (left) and the average profile (right). The orange lines correspond to our model, and blue to the \sersic profile. The dashed grey line represents the EWS noise level. Our simulations show more diverse profiles, but the average closely matches  the analytic expectations. The irregularities at very low S/N  on the FVAE profiles are a sign that the model does not produce perfectly noise-free galaxies. }
    \label{fig:light profiles}
\end{figure*}

We compare the radial profiles of generated galaxies with the  profiles of analytic galaxies with the same global properties.
Figure~\ref{fig:light profiles} compares and shows the radial profile for three bulge components, disk components, and the combination of the two components, simulated with our model and with \galsim. All the images are convolved by the VIS PSF but are without noise. We show both the profile along the major axis and the azimuthally averaged profile. The former is useful to identify deviations from a smooth profile, and thus highlights where the irregularities take place. The latter, computed by averaging the luminosity at a given radius $r$ from the galaxy centre in all directions, allows us to check if the average profile behaves as expected compared to the \sersic model. Overall, the figure shows the expected behaviour. Some profiles deviate significantly from a \sersic profile along the major axis. An example for this is the disk component shown in the bottom row of Fig.~\ref{fig:light profiles}, where we can see a spiral arm feature that creates variation in the radial light profile. However, the average profiles tend  to follow the analytic expectations since irregularities are averaged out. Therefore, the generated galaxies seem to present the desired behaviour (i.e. complex surface brightness distributions), which on average match a \sersic profile. An additional interesting result seen in  Fig.~\ref{fig:light profiles} is that the combination of the two components also behaves very similarly when compared to a double-component analytic galaxy (see the composite galaxy column).

\subsubsection{Surface brightness fitting}

We now fit \sersic models to quantify how accurately the shape parameters are recovered in a statistical sense. For this purpose we use the \texttt{Galapagos} package~\citep{galapagos,galapagos2}. \texttt{Galapagos} is a high-level wrapper for \texttt{SExtractor}~\citep{sextractor} and \texttt{Galfit}~\citep{galfit} to automatically fit large samples of galaxies. Because two-component \sersic fits are generally less stable than one-component fits (e.g. \citealp{Simard}, \citealp{bernardi}, \citealp{candels}) we decide to produce the two components separately in two distinct realisation of the field. Thus, we have two different fields, one with only the bulge component and one with only the disk component. We then fit each field with the  one-component \sersic model. This allows us to test the reliability of the fits while reducing the degeneracies. Since our objective is to compare our simulation to an analytic one, a single \sersic fit is enough for our purpose.

Using the Euclid Flagship catalogue, we generate with our model a galaxy field of $0.4\,\rm{deg}^2$ (i.e. around $2500$ galaxies with magnitude lower than $25$), following the same procedure as in Sect.~\ref{sec:large_field}. We then use the same procedure and subsample to produce the same field with the pure \sersic profiles. The two fields are therefore identical in terms of number of galaxies and  positions, and contain galaxies with the same structural properties.

\begin{figure}
    \centering
    \includegraphics[width = \linewidth]{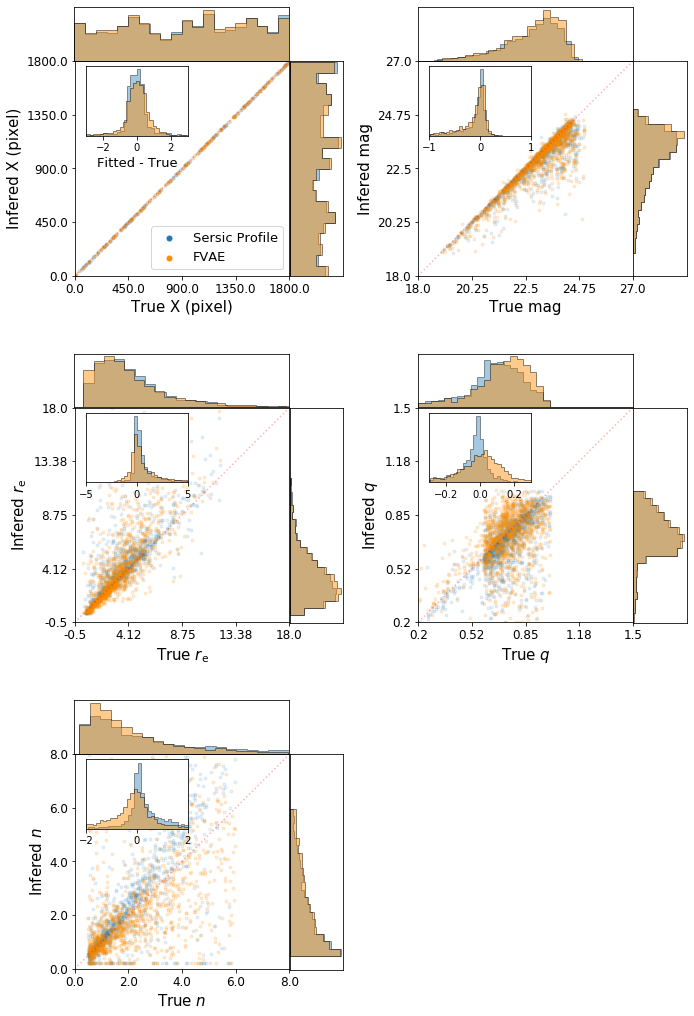}
    \caption{Results of 2D \sersic fits to the surface brightness distributions of bulge components. In every panel, the orange points and histograms represent the results for the FVAE galaxies and the blue for the analytic galaxies. Each panel represents a different parameter, as labelled. For each parameter  the true value of the parameter is plotted as a function of the inferred one from the best-fit model. A perfect fit corresponds to the diagonal. In addition, above and to the right of each plot are the projected distributions of the scatter plot. Finally, the inset plot shows the distribution of the error (fitted value minus true value). To make the scatter plot less crowded, only half the galaxies are plotted, but the error histograms and the projected distributions are computed on the entire field (for more details of the error distributions, see Table~\ref{tab:error_fit}).}
    \label{fig:fit_bulges}
\end{figure}

\begin{figure}
    \centering
    \includegraphics[width = \linewidth]{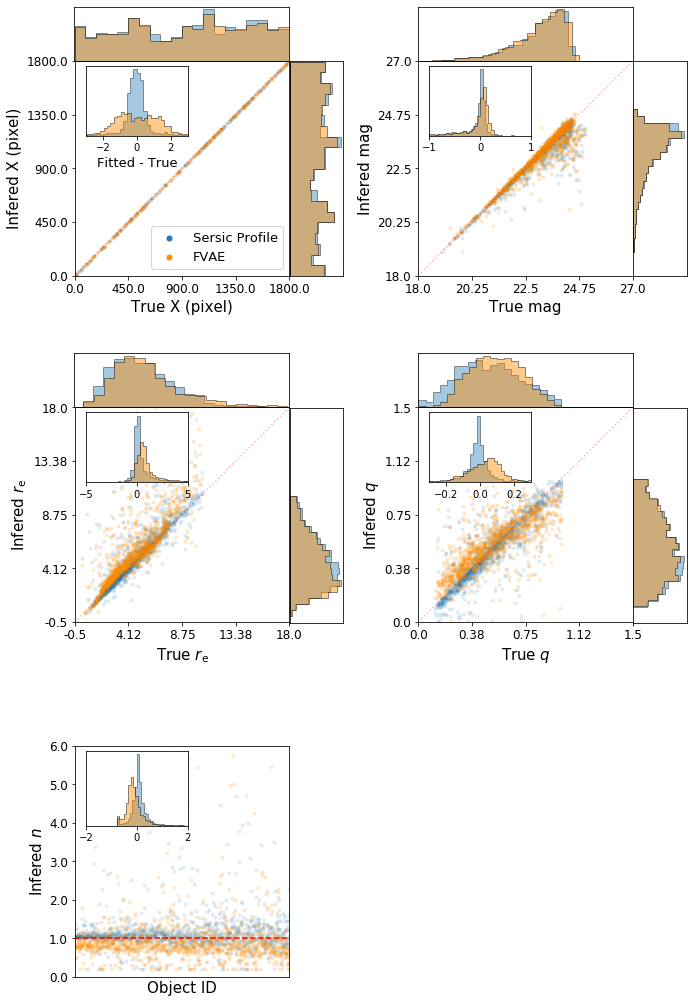}
    \caption{Same  as Fig.~\ref{fig:fit_bulges}, but for the results and description of the  disk component.}
    \label{fig:fit_disks}
\end{figure}

Figures \ref{fig:fit_bulges} and \ref{fig:fit_disks}  show the fitting results concerning  bulge and disk components, respectively, for five parameters: half-light radius $r_{\rm{e}}$, axis ratio $q$,  \sersic index $n$,  centroid position $X,$ and  total magnitude. We recall that the goal of this comparison is not to quantify the absolute accuracy of the fits, but to compare the relative behaviour of our simulations with a baseline. A future publication in preparation will quantify in detail the accuracy of structural parameters in both the EWS and the EDS. Overall, the structural parameters are recovered with similar dispersion for  the FVAE and the analytic simulation. This is a first quantitative confirmation of the visual inspection of the previous sections. Our model is able to produce realistic galaxy images while preserving information on the parametric structure. The global distributions of the predicted parameters are also very similar, which confirms that our model has correctly learnt the entire distribution of the training set, and is thus able to span the entire parameter space of the Euclid Flagship catalogue.

 Looking in more detail, the FVAE results present a slightly larger dispersion in all recovered parameters. This is expected since the analytic simulations represent a perfect match for the model that is fitted. This is not the case for the FVAE simulations, which present more complex profiles. 
 We give the statistical details of the fitting distribution errors (median, first, and third quartile) in Table~\ref{tab:error_fit}, corresponding to the distributions in the insets  in each panel of  Figs. \ref{fig:fit_bulges} and \ref{fig:fit_disks}.

\begin{table*}
\caption{Accuracy of fitting results. For each parameter shown in Figs.~\ref{fig:fit_bulges} (bulges) and \ref{fig:fit_disks} (disks), we present the first quartile ($q_1$), the median ($\mu_{1/2}$), and the third quartile ($q_3$) of the fitting error distributions.}
\centering
\begin{tabular}{|c|c|c|c||c|c|c|} 
\cline{2-7}
\multicolumn{1}{c|}{}    & \multicolumn{1}{c}{}   & \multicolumn{1}{c}{Bulges} &                        & \multicolumn{1}{c}{}   & \multicolumn{1}{c}{Disks} &                         \\ 
\hline
\diagbox{\textcolor{b}{Analytic}}{\textcolor{o}{FVAE}} & $q_1$                     & $\mu_{1/2}$                         & $q_3$                     & $q_1$                     & $\mu_{1/2}$                        & $q_3$                      \\ 
\hline
$X$                      & \diagbox{\textcolor{b}{$-0.37$}}{\textcolor{o}{$-0.36$}} & \diagbox{\textcolor{b}{$-0.04$}}{\textcolor{o}{$0.01$}}     & \diagbox{\textcolor{b}{$0.26$}}{\textcolor{o}{$0.40$}} & \diagbox{\textcolor{b}{$-0.36$}}{\textcolor{o}{$-0.90$}} & \diagbox{\textcolor{b}{$-0.05$}}{\textcolor{o}{$-0.01$}}    & \diagbox{\textcolor{b}{$0.25$}}{\textcolor{o}{$1.00$}}  \\ 
\hline
mag                        & \diagbox{\textcolor{b}{$-0.25$}}{\textcolor{o}{$-0.33$}} & \diagbox{\textcolor{b}{$-0.04$}}{\textcolor{o}{$-0.06$}}     & \diagbox{\textcolor{b}{$0.03$}}{\textcolor{o}{$0.03$}} & \diagbox{\textcolor{b}{$-0.11$}}{\textcolor{o}{$-0.09$}} & \diagbox{\textcolor{b}{$\,\,\,0.01$}}{\textcolor{o}{$\,\,\,0.04$}}   & \diagbox{\textcolor{b}{$0.04$}}{\textcolor{o}{$0.10$}}  \\ 
\hline
$r_{\rm{e}}$                & \diagbox{\textcolor{b}{$-0.04$}}{\textcolor{o}{$-0.23$}} & \diagbox{\textcolor{b}{$\,\,\,0.24$}}{\textcolor{o}{$\,\,\,\,0.25$}} & \diagbox{\textcolor{b}{$1.25$}}{\textcolor{o}{$1.77$}} & \diagbox{\textcolor{b}{$-0.07$}}{\textcolor{o}{$0.27$}} & \diagbox{\textcolor{b}{$\,\,\,0.11$}}{\textcolor{o}{$\,\,\,0.65$}}    & \diagbox{\textcolor{b}{$0.55$}}{\textcolor{o}{$1.27$}}  \\ 
\hline
$q$                        & \diagbox{\textcolor{b}{$-0.10$}}{\textcolor{o}{$-0.10$}} & \diagbox{\textcolor{b}{$-0.03$}}{\textcolor{o}{$0.00$}}     & \diagbox{\textcolor{b}{$0.00$}}{\textcolor{o}{$0.07$}} & \diagbox{\textcolor{b}{$-0.05$}}{\textcolor{o}{$-0.04$}} & \diagbox{\textcolor{b}{$-0.01$}}{\textcolor{o}{$0.03$}}   & \diagbox{\textcolor{b}{$0.01$}}{\textcolor{o}{$0.09$}} \\ 
\hline
$n$                        & \diagbox{\textcolor{b}{$-0.01$}}{\textcolor{o}{$-0.64$}} & \diagbox{\textcolor{b}{$\,\,\,0.23$}}{\textcolor{o}{$-0.06$}}     & \diagbox{\textcolor{b}{$1.26$}}{\textcolor{o}{$0.52$}} & \diagbox{\textcolor{b}{$-0.06$}}{\textcolor{o}{$-0.29$}} & \diagbox{\textcolor{b}{$0.06$}}{\textcolor{o}{$-0.15$}}   & \diagbox{\textcolor{b}{$0.20$}}{\textcolor{o}{$0.04$}}  \\
\hline
\end{tabular}

\label{tab:error_fit}
\end{table*}

The systematic offsets might be more problematic. The figure shows that the systematic shifts for the bulge components are very similar for the analytic and the FVAE fields which means that using a FVAE does not introduce any noticeable systematic effects. The only parameter that presents a small bias towards larger values is the axis ratio $q$.  This might be because of a lack of very elongated bulges in the training data set. The disk components present a slightly higher systematic bias though, as shown in Fig.~\ref{fig:fit_disks}. Indeed, FVAE galaxies tend to be systematically larger and rounder than their analytical counterparts and show an almost constant offset of $0.15$ on the \sersic index. It is not obvious whether these offsets are a consequence of the simulation or whether it is related to the fitting procedure itself. A possible explanation for the larger offsets is that  disk components are generally more extended and with flatter profiles than bulge components,  thus they also present more complexity and structure. Alternatively, it can also be related to the simulation itself. Our training set is   based on a single-component fit with a continuous distribution of the \sersic index. However, the \sersic index of the disk component in the Euclid Flagship catalogue is fixed to $n=1$. This means that there is only a small number of examples in the training set with exactly $n=1$, which can affect the quality of the generation.
Finally, we can see that for the magnitude, the fit of our galaxies also differs  very little from the \sersic fits, even if the flux is not something that is parametrised in our model, but re-scaled afterwards with \texttt{Galsim}. This occurs because the recovering of the flux in a large field, with blended galaxies for example, is not completely trivial.

\section{Forecasts for galaxy morphology with \Euclid}\label{sec:forecast}
The previous sections have shown that our proposed framework successfully generates galaxies with realistic and resolved structure. Our simulations can therefore be used to establish some forecasts on the number of galaxies for which \Euclid will be able to resolve the internal structure beyond a \sersic profile. 

\subsection{Identifying galaxies with resolved structure}

\begin{figure*}[h]
    \centering
    \includegraphics[width = \linewidth]{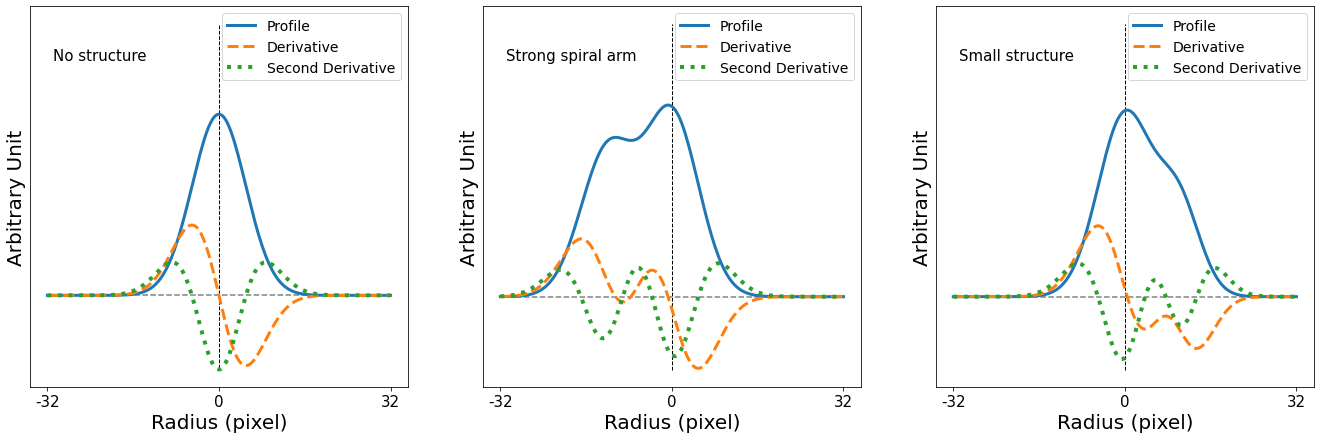}
    \caption{Three  toy profiles that illustrate our structure detection method. The left panel shows a smooth galaxy without structure, the middle panel a strongly perturbed galaxy, and the right panel a slightly perturbed object. For each profile,   its luminosity is plotted as a function of the distance to the galaxy centre in arbitrary units (blue solid lines). Their corresponding first and second derivatives are also plotted (orange and green solid lines, respectively). We can see that the number of roots in the second derivative is a good indicator of perturbed galaxies.}
    \label{fig:structure_method}
\end{figure*}

Our goal is to quantify the fraction of galaxies that present significant structures that deviate from a pure analytic profile. For that purpose we have designed a method to distinguish galaxies with internal structure from smooth objects. We assume that any type of complexity in the galaxy surface brightness distribution, hereafter called structure, will result in a deviation from an analytical profile. This is particularly clear in the disk component shown in Fig.~\ref{fig:light profiles}. We therefore establish a criterion to characterise the smoothness of a galaxy by computing the derivatives of the semi-major axis profile. For illustration purposes, we show in Fig.~\ref{fig:structure_method} three toy profiles. A pure analytical profile, a profile presenting a strong structure, and  a slightly perturbed one. When the profile is smooth the first derivative is also smooth, changing its sign only at the centre of the galaxy. If we consider only a one-sided profile, the derivative never goes to zero (i.e. it has no roots). Its second derivative is also smooth, and has only one root that we call a `natural zero'. When the galaxy is strongly perturbed, the profile will significantly differ from a pure analytical profile. For a \sersic profile the light curve   decreases from the  centre to the edge of the galaxy; instead, for example in a galaxy presenting a spiral arm, the major axis profile increases in the location of the arm. This increase (change of slope) will cause a sign change in the first derivative, and thus two changes in sign in the second derivative, as can be seen in the second column of Fig.~\ref{fig:structure_method}. However, the roots of the first derivative are not always enough to detect a variation from a smooth profile, as illustrated in the third column of the figure; the profile can be slightly perturbed, with a change of slope in the profile, but this does not make the profile rise as in the second column of the figure, but significantly changes  the rate of  decrease. Thus, the first derivative will not change in sign (the profile does not increase), but the second derivative will (the rate of the decrease changes).

Therefore, we conclude that the presence of a zero on the second derivative of the light profile (without counting the natural zero) is a reasonable indicator of a galaxy with complex structures. We note that there will be additional zeros at the edge of the profile when it becomes flat. However, these roots will be all consecutive, giving us a way to distinguish zeros coming from a structure from ones coming from the end of the profile. Thus, we can consider a galaxy being structured if its second derivative has two roots (without considering the first natural one), which are far enough from each other. This also prevents the high-frequency perturbations in the profile that we do not want to consider as a structure. We find that, at the VIS resolution, a minimum distance of $1\,\rm{pixel}$ (approximately one PSF FWHM) between roots is a reasonable choice. To make sure that we do not miss structures that are not along the semi-major axis, we also search for structures with the same method along the semi-minor axis of the galaxy.

We show in Appendix \ref{sec:annexe1} two random selections of galaxies which have been classified with and without structure. Our method successfully isolates galaxies with perturbed or asymmetric profiles.

\subsection{Resolved complex morphologies in \Euclid}

\begin{figure}
    \centering
    \includegraphics[width = \linewidth]{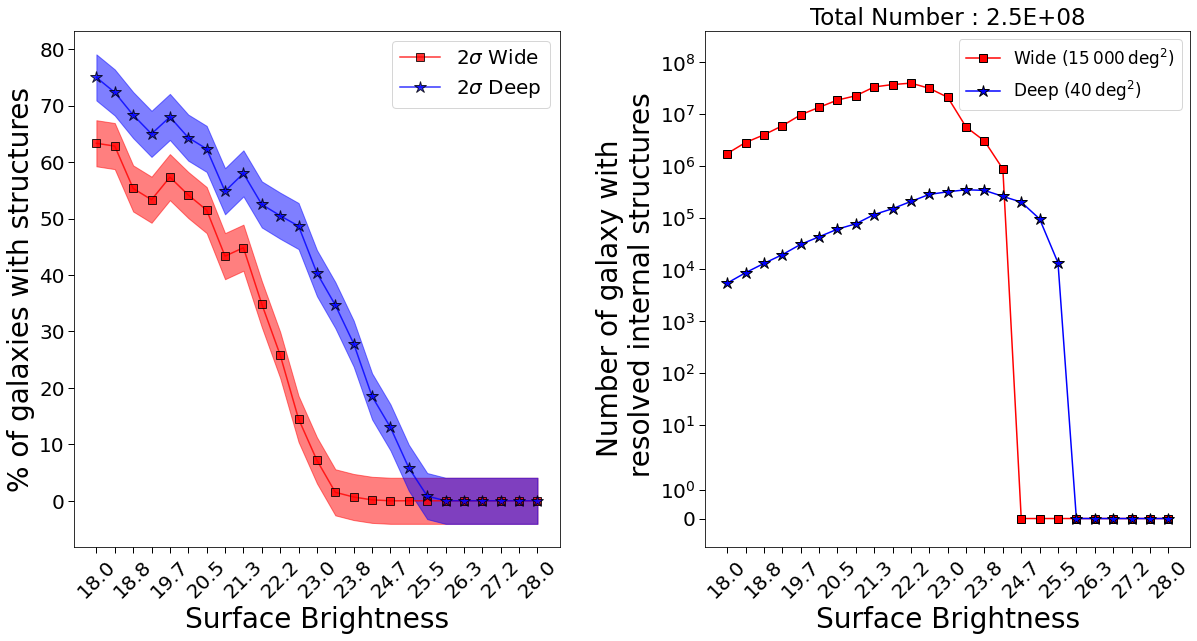}
    \caption{Forecast of the number of galaxies with internal structures for the EWS and the EDS, regarding surface brightness. Left panel: Fraction of galaxies with resolved structure as a function of surface brightness. Right panel: Total number of galaxies with resolved structure as a function of surface brightness. The red squares are for structures discernible for the EWS at $2\,\sigma$ around the noise level. The blue stars represent the same information, but for the EDS.}
    \label{fig:percentage_SB}
\end{figure}

We use this technique  to infer the fraction of galaxies for which \Euclid will be able to resolve internal morphological structure beyond \sersic profiles. We simulate galaxies without noise and compute the semi-major axis profile and consider only pixels $2\,\sigma$ above the noise level. We then plot in the left plot of Fig.~\ref{fig:percentage_SB} the fraction of galaxies presenting structures as a function of the surface brightness $S_{\rm{b}}$ of the galaxy, defined as
\begin{equation}
    S_{\rm{b}} = m + 2.5\log_{10}(\pi\, q_{\rm{tot}}\,r_{\rm{tot}}^2)\,,
\end{equation}
where $r_{\rm{tot}}$ (in arcsecond) and $q_{\rm{tot}}$ are the global (disk and bulge components) half-light radius and axis ratio of the galaxy. Thus, $\pi\,q_{\rm{tot}}\,r_{\rm{tot}}^2$ represents the area of the galaxy.

We can see that the fraction of galaxies with resolved structures decreases with increasing surface brightness, as expected. The behaviour of the EWS and the EDS is self-similar, but the EDS is shifted towards fainter surface brightness. The difference is of the order of 2 magnitudes: less than $10\,\%$ of galaxies present detailed structures above $2\,\sigma$, beyond a surface brightness of $22.5$ $\rm{mag}\,\rm{arcsec}^{-2}$ for the EWS and $24.9\,\rm{mag}\,\rm{arcsec}^{-2}$ for the EDS. The statistical fluctuations on the curve are similar because we compute our structure indicator on the same realisations of galaxies with only the S/N changing.

We also provide the total number of galaxies per bin in the right panel of Fig.~\ref{fig:percentage_SB}. We simply multiply the fraction of objects with structure by the total number of galaxies in the $15\,000\,\rm{deg}^2$ of the EWS and in the $40\,\rm{deg}^2$ of the EDS. We conclude that \Euclid will observe around $250\,\rm{million}$ galaxies that are significantly more complex than the analytical profiles during the six years of the mission.

\begin{figure}
    \centering
    \includegraphics[width = \linewidth]{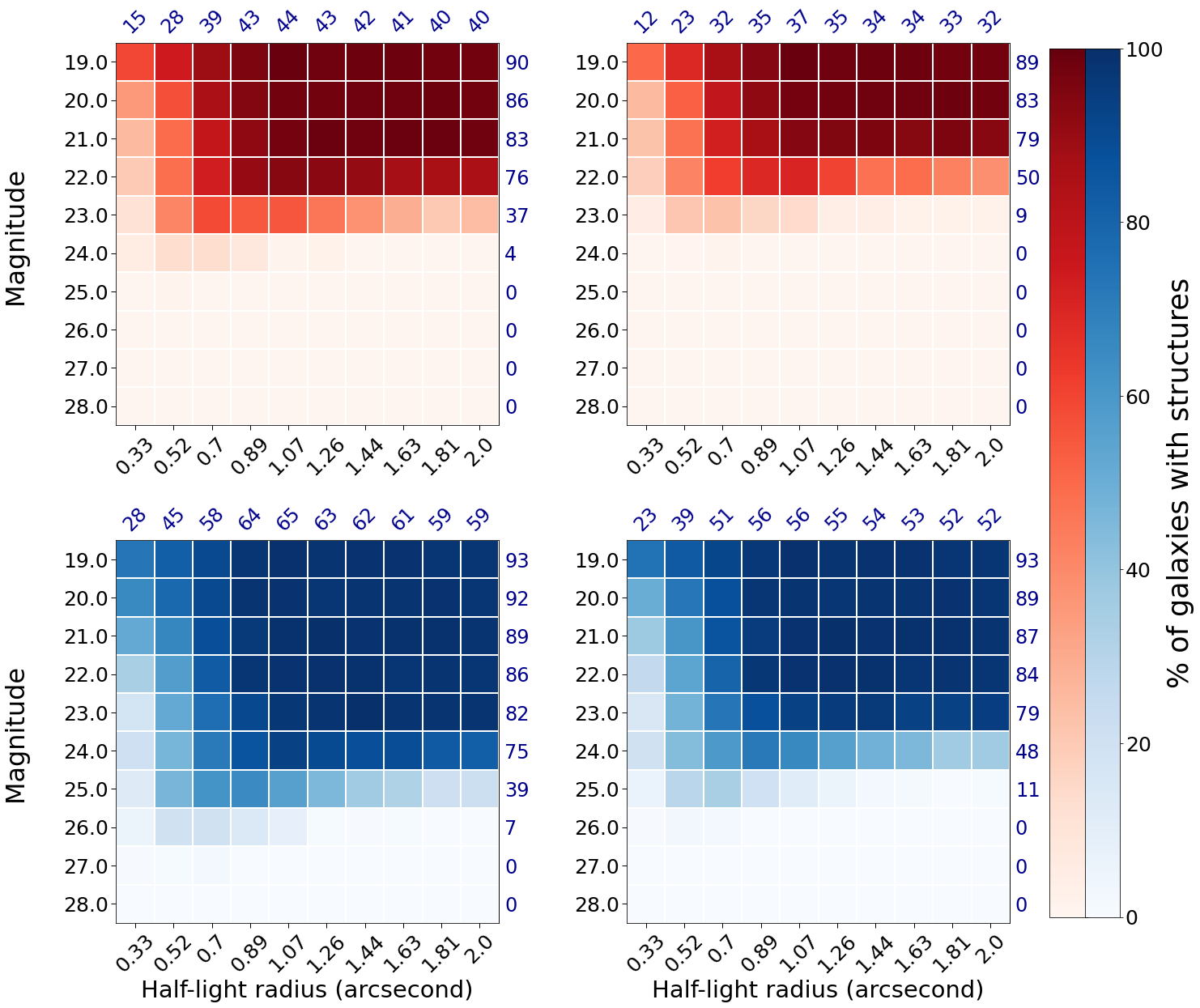}
    \caption{Fraction of galaxies with resolved structures in bins of magnitude and half-light radius. The first line represents \Euclid capacities for the EWS, and the second for the EDS. The first column is the percentage of galaxies presenting structures above $1\,\sigma$ of the noise level, and the second column above $2\,\sigma$. The colour-coding is the same as in Fig.~\ref{fig:percentage_SB} and \ref{fig:forecast_logmz}. The blue number in each column (row) indicate the mean percentage of the corresponding column (row).}
    \label{fig:percentage_grid}
\end{figure}

 Figure~\ref{fig:percentage_grid} shows a 2D representation of the fraction of galaxies with resolved structures above $1\,\sigma$ and $2\,\sigma$ of the noise as a function of magnitude and half-light radius. We observe the same behaviour, namely that the EDS goes around two magnitudes deeper to probe morphologies. The fraction of galaxies decreases in the limit of the distributions when we increase the level of acceptance from $1\,\sigma$ to $2\,\sigma$. The figure summarises the following expected behaviour: (1) the brighter the galaxy, the larger  the number of resolved structures (top to bottom gradient) and (2) the fraction becomes smaller for extremes (very small and very large galaxies) at constant magnitude. The decrease at small sizes is a consequence of resolution. At large sizes it is related to  S/N. We recall that we did not plot galaxies bigger than $2\,\arcsecond$ because of the built-in size limitation of our model, but we expect the decreasing trend to continue at larger radii.

\begin{figure}
    \centering
    \includegraphics[width = \linewidth]{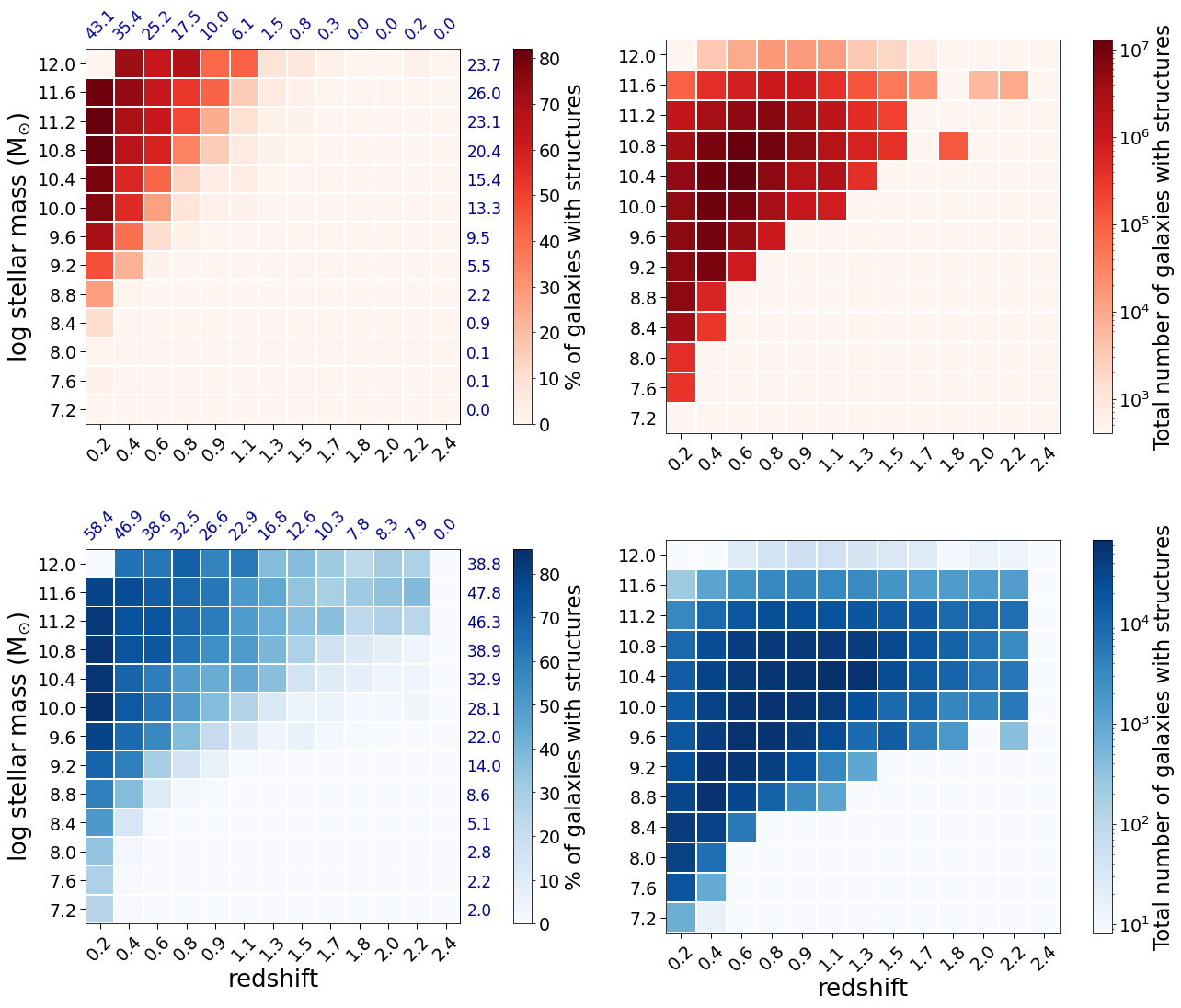}
    \caption{Fraction of galaxies with resolved structures in bins of stellar mass and redshift. The top (bottom) row corresponds to the EWS (EDS). The left column indicates fractions, while the right column are absolute numbers for the six-year survey. The numbers in blue indicate the average values per row and column. }
    \label{fig:forecast_logmz}
\end{figure}

Finally, in Fig.~\ref{fig:forecast_logmz} we forecast the fraction and the total number of galaxies with resolved structures as a function of physical properties of galaxies, namely stellar mass and redshift. We conclude that the EWS will be able to reach a $50\,\%$ completeness regarding the detection of internal structures of galaxies down to $\sim10^{10.6}\,\rm{M}_\odot$ at $z\sim0.5$. The EDS reaches down to a stellar mass of $10^{9.6}\,\rm{M}_\odot$ up to $z\sim0.5$.

We note here that we are probing the internal structures of the galaxies, and not assessing whether the galaxy is resolved or not.  We thus consider, in our forecasts, that intrinsically smooth galaxies such as spheroids have no structures, even if they are resolved by \Euclid. Since our model is trained on real data, it is reasonable to assume that the fraction of different morphologies is well reproduced. The numbers we provide are therefore an estimate of the fraction of galaxies with complex internal structure, beyond a \sersic model.

\section{Discussion: A framework to simulate future surveys}\label{sec:discussion}

This work presents a novel framework to generate galaxies with realistic morphologies, while keeping  control on the global structural properties. It can be used to calibrate algorithms for future experiments such as \Euclid in which the impact of complex galaxy shapes might become significant. This is the case for example for galaxy deblending or even shear estimations. We discuss in this section possible limitations of a large-scale use of generative models for galaxy generation.

\begin{figure}
    \centering
    \includegraphics[width = \linewidth]{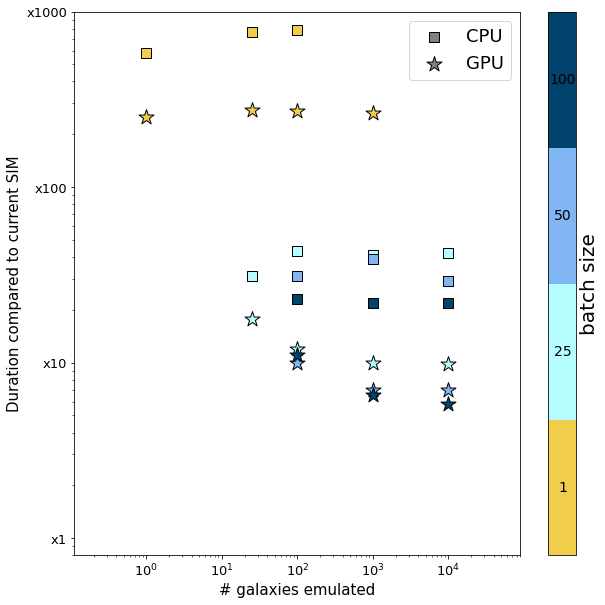}
    \caption{Comparison of the execution time between our model and the current \Euclid simulations, for different hardware configurations. The $y$-axis indicates the ratio of the execution time using our model to the time from the official \Euclid pipeline.The $x$-axis corresponds to the number of galaxies simulated. The stars represent GPU runs and squares are CPU runs. The colour bar indicates the batch size.}
    \label{fig:speed}
\end{figure}

One possible bottleneck is execution time. We therefore quantify the execution speed of our framework  compared to that of  a classical analytic generation. We use two different environments: with and without GPUs. We used a $16$ CPU Intel Xeon Bronze $3106$, and an NVIDIA Tesla P$40$ GPU.
We then tested our method with increasing batch sizes, going from one galaxy at a time to $64$. The results of the different experiments are summarised in Fig.~\ref{fig:speed}. Each measurement  refers to the execution time of a standard analytic simulation. The training time is not discussed here as it has to be done only once, and does not enter execution time discussions.

The figure confirms that a GPU is around four times faster than a CPU environment in all configurations. We also see that the batch size has a dramatic impact on the execution speed. For a batch size of one, our method is more than a $100$ times slower than a traditional approach. However, the difference is reduced to a factor of around $5$ if larger batches are used. It is interesting to note, however,  that the execution time does not depend linearly on the batch size. This is a well-known behaviour~\citep{latency}. 
We note that for this figure, as in all this work, we simulate galaxies by a sum of two components. As explained before, we did that to match the current \Euclid simulation strategy. Nevertheless, we highlight here that we are capable of creating complex galaxies with only one component, and therefore all the times of the figure could possibly be divided by two if we were simulating only one component.

Another possible limitation of our proposed framework is the fact that it is trained on observations which therefore contain biases that can propagate the simulation. In particular, we used here the \cosmos \texttt{Galfit} fitting as a ground truth to condition the autoregressive flow. The impact of this could be assessed by using different independent fitting codes on the same data sets and comparing the results. This is an ongoing effort as part of the \Euclid Morphology Challenge, which will be presented in a forthcoming publication. The diversity of generated galaxies is also limited by the quality of the observations used for training, in this case HST observations. This restricts the range of parameters that our model can probe without extrapolation. This could be mitigated with additional data sets, but we have not explored that in this work. For this reason, we do  not recommend  using our framework to simulate images without noise as features below the HST noise level are not constrained. We also note that our model is limited regarding the size of the galaxies it can generate because of the fixed size of the training stamps. Simulating galaxies with half-light radii bigger than $2\arcsecond$ is not recommended. Some galaxies larger than $\ang{;;1.5}$ and with a small \sersic index (flux above the half-light radius not negligible) can also produce some flux artefacts at the border of the stamp, being at the limit of the training distribution, and because the faint end of the galaxy will be cut. We used this large limit of $2\arcsec$ to do our morphological forecasts because those artefacts do not cause problems in our structure detection algorithm.
In addition, to produce galaxies fainter than the limiting magnitude of the training set ($25.2\,\rm{mag}$), we assumed that the galaxy morphology is not correlated with the magnitude, which is of course an approximation. 
Finally, to establish morphology forecasts, we assume that the amount of structures produced by our model is the same as in real galaxies. Our model may tend to produce galaxies that are smoother than in real galaxies. Therefore our forecast may have underestimated the number of objects with complex morphologies. On the contrary, our choice to use fields without any instrumental effects but the PSF could decrease the effective number of detected galaxies. Finally, the number of low-magnitude galaxies in the Euclid Flagship catalogue could be underestimated, for example compared to the catalogue \citep{lsst_cat} used for the Legacy Survey of Space and Time (LSST) at the Vera C. Rubin Observatory \citep{lsst}. This lack of faint galaxies could increase the numbers presented here, especially for the EDS.

\section{Summary and conclusion}\label{sec:summary}
We have presented a data-driven method for simulating deconvolved and noise-free galaxies with morphologies more realistic and complex than pure analytic \sersic profiles. The proposed approach is based on a combination of deep generative neural networks trained on observations, which allows one to generate realistic galaxies while preserving a control of the global shape of the surface brightness profiles. We have shown that the structural parameters of the generated galaxies are recovered with similar accuracy compared to that derived for analytic profiles. Our proposed approach, although around five times slower than an analytic simulation can be used to generate realistic simulations for future missions and  experiments, and therefore calibrate algorithms under more realistic conditions.

We have used this new framework to establish the first forecasts on the number of galaxies for which \Euclid will be able to provide resolved morphological structure beyond \sersic profiles. We find that \Euclid will resolve the internal structure of around $250\,\rm{million}$ galaxies. This corresponds to a $50\,\%$ stellar mass complete sample above $10^{10.6}$ ($10^{9.6}$) at a redshift $z=0.5$ for the EWS (EDS). This is a first estimation of the capabilities of \Euclid for estimating galaxy morphologies, which are a key ingredient for a variety of galaxy evolution-related science cases.

Looking ahead, there is an ongoing effort of the authors to adapt the VAE to work in a multi-band mode, which will enable the generation of galaxies in the two infrared bands of the Euclid near-infrared imager. We also plan to train a flow on different sets of parameters. Our method can, for example, be conditioned on the orientation and the environment of galaxies to take into account gravitational shear effects. We could also condition our flow on the redshift and initial mass function in order to find their impact on the evolution of structures.

\begin{acknowledgements}
We thank the IAC where the first author was in long term visit during the production of this paper, with a special thanks to the \textit{TRACES} team for their support. We would also like to thank the Direction Informatique de l'Observatoire (DIO) of the Paris Meudon Observatory for the management and support of the GPU we used to train our deep learning models. We also thank the Centre National d'Etudes Spatiales (CNES) and the Centre National de la Recherche Scientifique (CNRS) for the financial support of the PhD in which this study took place.
This work has made use of CosmoHub.

CosmoHub has been developed by the Port d'Informació Científica (PIC), maintained through a collaboration of the Institut de Física d'Altes Energies (IFAE) and the Centro de Investigaciones Energéticas, Medioambientales y Tecnológicas (CIEMAT) and the Institute of Space Sciences (CSIC \& IEEC), and was partially funded by the "Plan Estatal de Investigación Científica y Técnica y de Innovación" program of the Spanish government. 
\AckEC
\\
\\
Softwares: Astropy \cite{astropy, astropy2}, \galsim \citep{galsim}, IPython \citep{ipython}, Jupyter \citep{jupyter}, Matplotlib \citep{plt}, Numpy \citep{numpy}, TensorFlow \citep{tf}, TensorFlow Probability \citep{tfp}.
\end{acknowledgements}

%
%

\clearpage

\bibliographystyle{aa.bst}
\bibliography{biblio}

\appendix
\section{More illustrations of the capabilities of our model.}\label{sec:annexe1}

\begin{figure}[!ht]
    \centering
    \includegraphics[width =\linewidth]{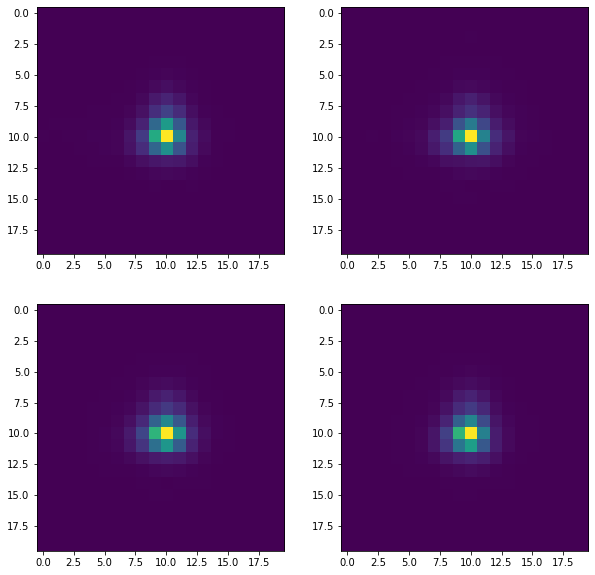}
    \caption{Simulation by our model of bulge components with radius smaller than $\ang{;;0.1}$ (i.e. smaller than one pixel). These bulge components are at the end of the Euclid Flagship radii distribution (Fig.~\ref{fig:data hists}) outside the \cosmos training domain. As can be seen,  our model is able to extrapolate from those cases. Because the object is not resolved, it is almost purely the VIS PSF.}
    \label{fig:tiny_bulges}
\end{figure}

\begin{figure}[!ht]
    \centering
    \includegraphics[width =\linewidth]{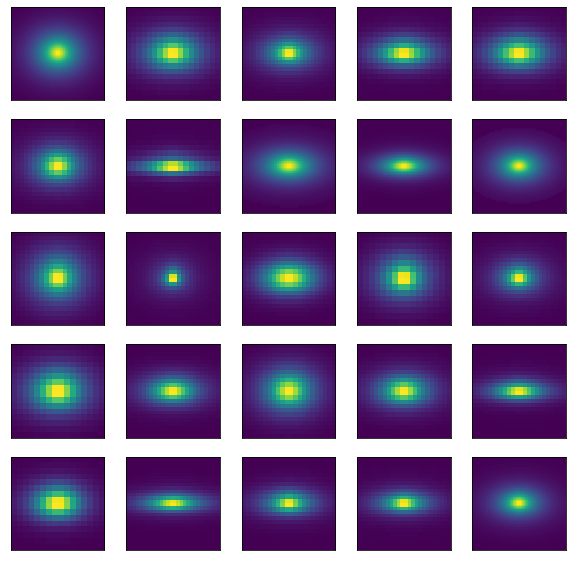}
    \caption{Double-\sersic component galaxies currently used in the Euclid Consortium.}
    \label{fig:sersic_gals}
\end{figure}

\begin{figure*}[!ht]
    \centering
    \includegraphics[scale=0.5]{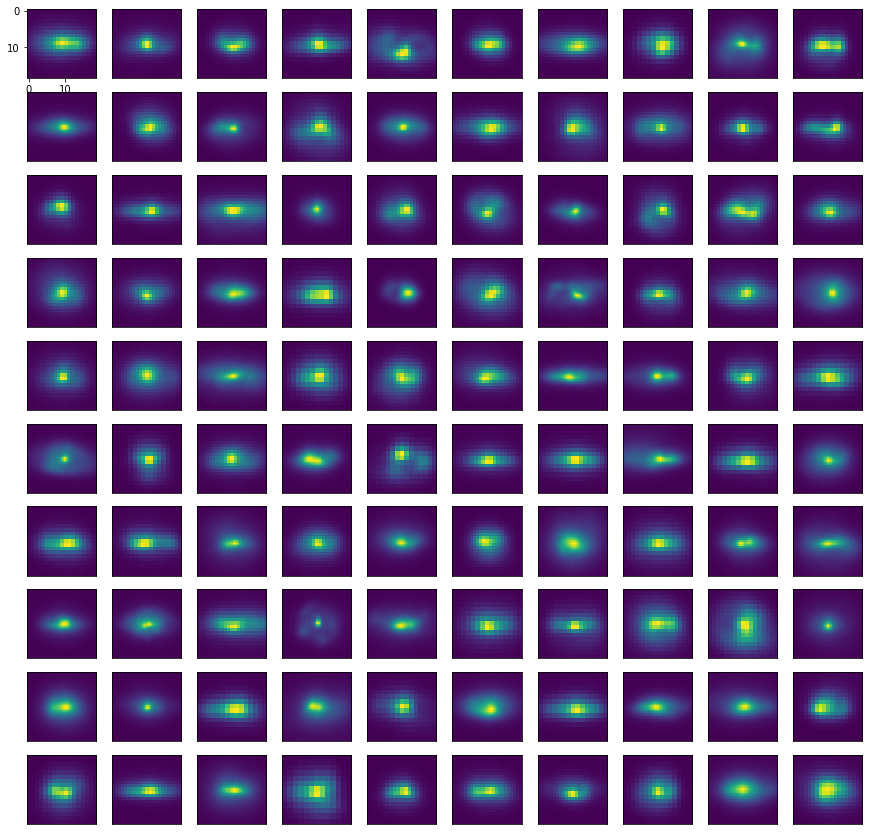}
    \caption{Random examples of galaxies considered as having structure. The stamps are cut at twice the effective radius.}
    \label{fig:structured_gals}
\end{figure*}

\begin{figure*}[!ht]
    \centering
    \includegraphics[scale=0.5]{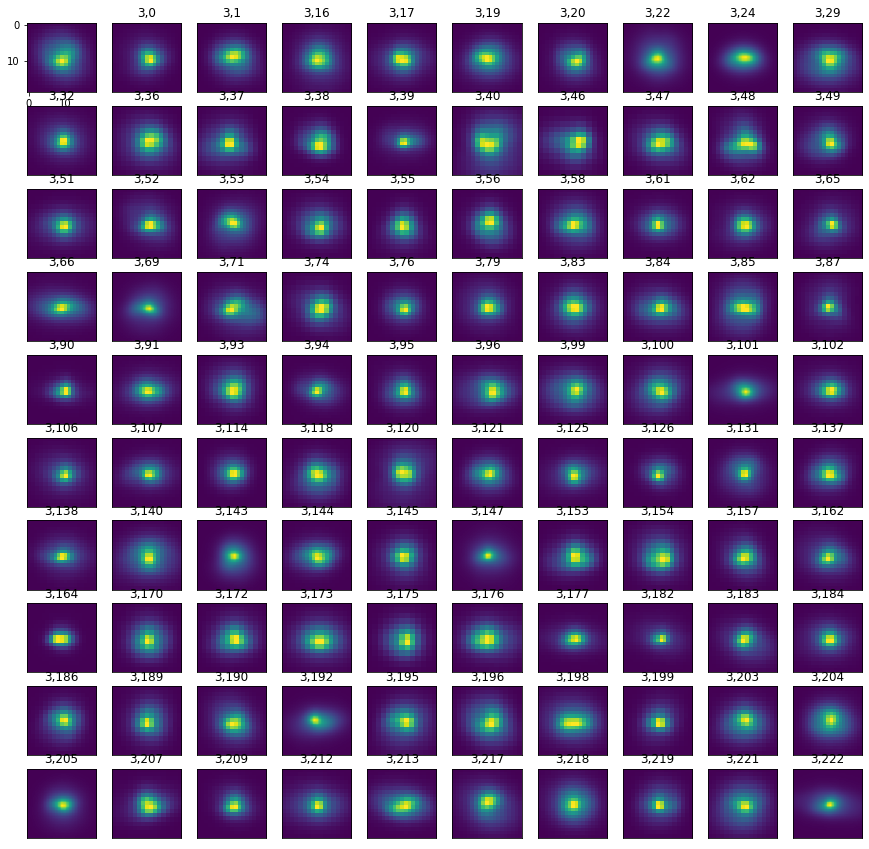}
    \caption{Random examples of galaxies considered as having no structure. The stamps are cut at twice the effective radius.}
    \label{fig:no_strctures}
\end{figure*}

\section{Detailed architectures of the FVAE.}

\begin{sidewaysfigure*}
    \centering
    \includegraphics[scale=0.4]{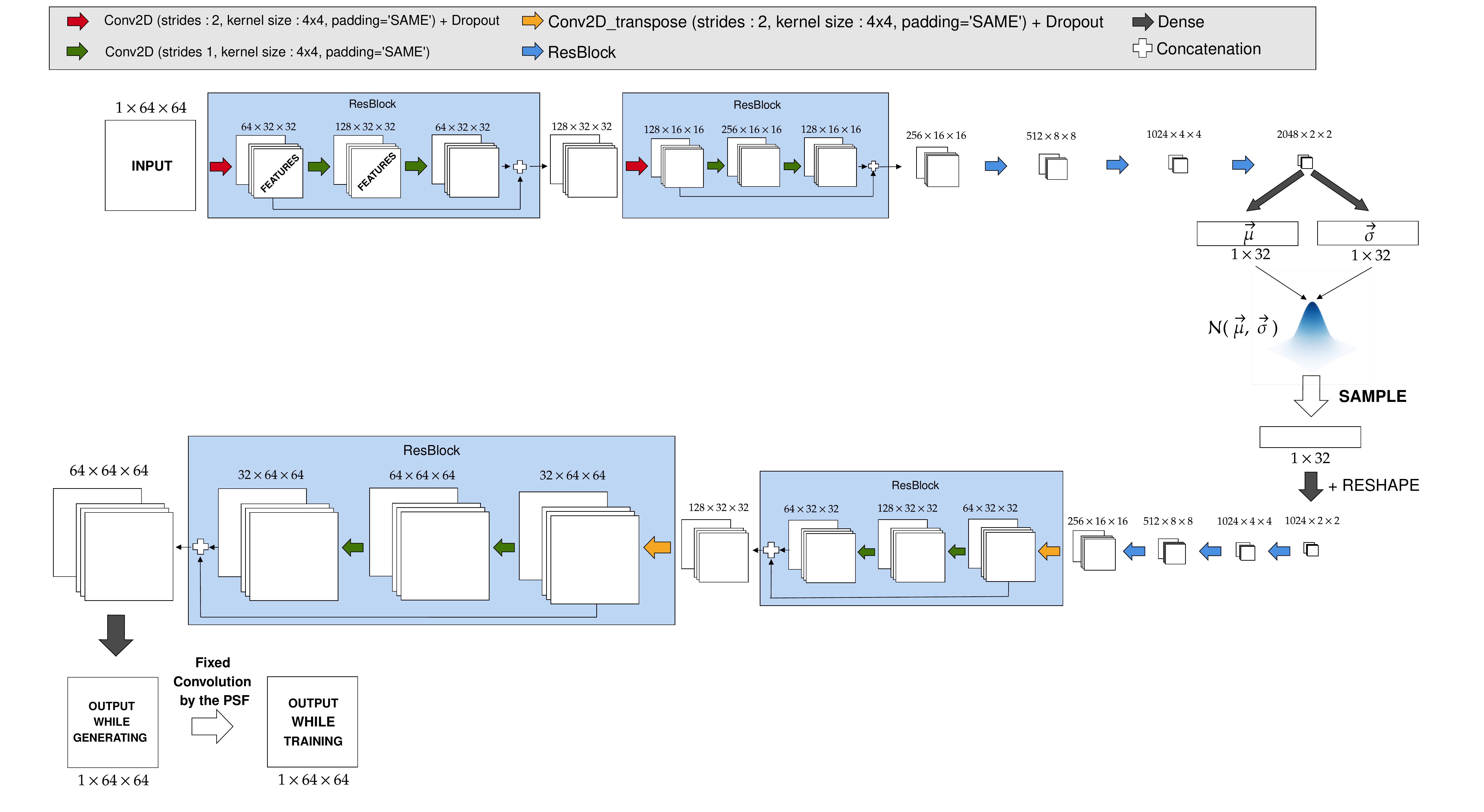}
    \caption{Schematic architecture of the VAE. Each convolution is followed by a ReLU activation and a dropout.}
    \label{fig:vae_architecture}
\end{sidewaysfigure*}

\begin{figure*}
    \centering
    \includegraphics[width = \linewidth]{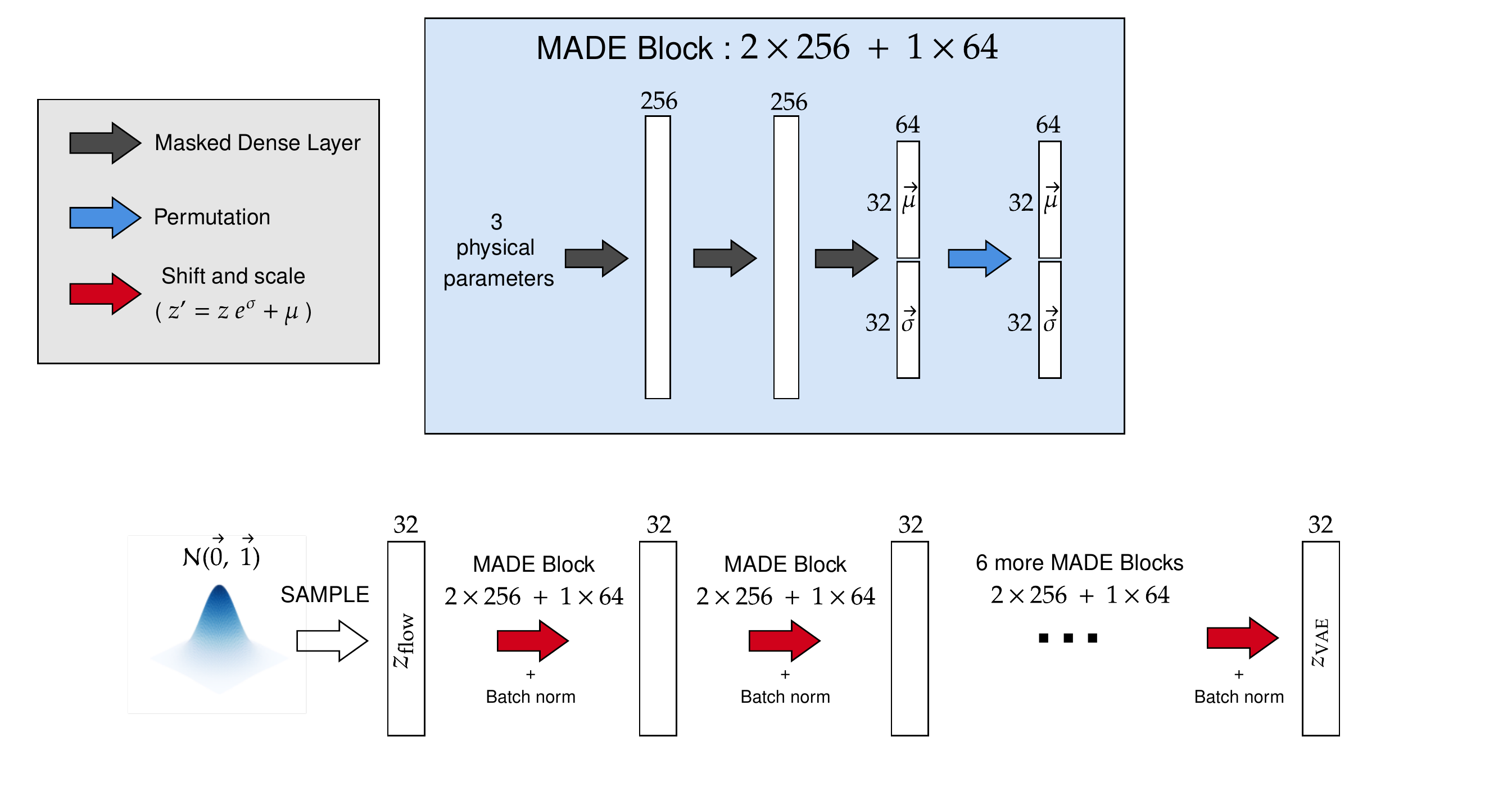}
    \caption{Schematic architecture of the regressive flow. Each dense layer is followed by a ReLU activation.}
    \label{fig:flow_architecture}
\end{figure*}

\end{document}